\documentclass[10pt,aps,prd,twocolumn,notitlepage,nofootinbib,floatfix,superscriptaddress,preprintnumbers]{revtex4-1}

\usepackage{amsmath, amsthm, amssymb, amsfonts, amsbsy, mathrsfs}
\usepackage[caption=false]{subfig}

\usepackage{xcolor}
\usepackage{calligra,bm}
\usepackage{stmaryrd}
\usepackage{subfig}

\usepackage{graphicx}
\graphicspath{{./images/}}

\usepackage[hidelinks,bookmarks=true]{hyperref} %make sure it comes last of your loaded packages
\hypersetup{pdfstartview=FitH,pdfhighlight=/O,colorlinks=false}

\begin{document}
\preprint{YITP-22-18, IPMU22-0004}

\title{Junction conditions and sharp gradients in generalized coupling theories.}

\author{Justin C. Feng}
\affiliation{CENTRA, Departamento de F{\'i}sica, Instituto Superior T{\'e}cnico (IST), \\ Universidade de Lisboa (UL), Avenida Rovisco Pais 1, 1049 Lisboa, Portugal}

\author{Shinji Mukohyama}
\affiliation{Center for Gravitational Physics, Yukawa Institute for Theoretical Physics,\\ Kyoto University, Kyoto 606-8502, Japan}
\affiliation{Kavli Institute for the Physics and Mathematics of the Universe (WPI), \\The University of Tokyo Institutes for Advanced Study,\\The University of Tokyo, Kashiwa, Chiba 277-8583, Japan}

\author{Sante Carloni}
\affiliation{DIME Sezione Metodi e Modelli Matematici, Universit\`{a} di Genova,\\ Via All'Opera Pia 15, 16145 Genoa, Italy.}

%\date{\today}

%-----------------------------------------------------------------------
%-----------------------------------
%-----------------
%--------
%---
%-
%
%
%-
%---
%--------
%-----------------
%-----------------------------------
%-----------------------------------------------------------------------

%=======================================================================
%-----------------------------------------------------------------------
%
%		ABSTRACT
%
%-----------------------------------------------------------------------
%=======================================================================
\begin{abstract}
 In this article, we develop the formalism for singular hypersurfaces and junction conditions in generalized coupling theories using a variational approach. We then employ this formalism to examine the behavior of sharp matter density gradients in generalized coupling theories. We find that such gradients do not necessarily lead to the pathologies present in other theories of gravity with auxiliary fields. A detailed example, based on a simple instance of a generalized coupling theory called the MEMe model, is also provided. In the static case, we show that sharp boundaries do not generate singularities in the dynamical frame despite the presence of an auxiliary field. Instead, in the case of a collapsing spherical density distribution with a general profile an additional force compresses over-densities and expands underdensities. These results can also be used to deduce additional constraints on the parameter of this model.
\end{abstract}

%\pacs{}
%\keywords{}

%-----------------------------------------------------------------------
%-----------------------------------
%-----------------
%--------
%---
%-
%
%
%-
%---
%--------
%-----------------
%-----------------------------------
%-----------------------------------------------------------------------

\maketitle

%-----------------------------------------------------------------------
%-----------------------------------
%-----------------
%--------
%---
%-
%
%
%-
%---
%--------
%-----------------
%-----------------------------------
%-----------------------------------------------------------------------

%=======================================================================
%-----------------------------------------------------------------------
%
%		INTRODUCTION
%
%-----------------------------------------------------------------------
%=======================================================================

\section{Introduction}
Generalized coupling theories (GCTs) are modified gravity theories characterized by a nontrivial coupling between matter fields and the spacetime metric $g_{\mu \nu}$ mediated by a coupling tensor \cite{Carloni:2016glo,Feng:2019dwu}. If this tensor is nondegenerate, one can construct a class of theories in which matter couples with an effective metric $\mathfrak{g}_{\mu \nu}$ while the coupling tensor is an auxiliary field satisfying a nontrivial algebraic field equation. In this way, one naturally distinguishes two frames,  an ``Einstein frame'' in which the basic variables are the spacetime metric and the coupling tensors, and a ``Jordan frame'' in which  $\mathfrak{g}_{\mu \nu}$ is the key variable.

A particularly simple, yet interesting, example of a GCT is the so-called minimal exponential measure (MEMe) model, which was proposed in \cite{Feng:2019dwu}, and can be recognized, together with a large class of GCTs, as a Type I minimally modified gravity (MMG) theory according to the classification scheme of \cite{Aoki:2018brq,DeFelice:2020eju,Aoki:2020oqc}. 

A relevant feature of the MEMe model is that near a certain critical density, the gravitational field equations simplify to the vacuum Einstein equation with a large cosmological constant. This feature is of particular interest in early universe cosmology and for compact objects. Regarding the former, one qualitatively has inflationary behavior in the early universe with a graceful exit (a feature shared with Eddington-Inspired Born-Infeld theories \cite{Banados:2010ix,BeltranJimenez:2017doy}). Regarding compact objects, the MEMe model may admit solutions describing gravastars \cite{Mazur:2001fv,Visser:2003ge,Mazur:2004fk} without the need for a false vacuum. Finally, in a recent paper \cite{Feng:2020lhp}, the parametrized Post-Newtonian (PPN) limit of the MEMe model was also studied, showing that GCT theories, like all the type-I MMG theories, require a special extension of the PPN formalism to be analyzed in the weak field approximation.

It has been known from some time \cite{Pani:2012qd} that auxiliary field theories coupled to gravity generally become pathological in the presence of sharp gradients of the energy-momentum tensor, which may occur e.g. at the surfaces of compact objects. In particular, it was argued that since auxiliary field theories generally contain derivatives of the energy-momentum tensor (more generally, sources of curvature), sharp gradients in the energy-momentum tensor can generate curvature singularities. Of course, this is not necessarily a fundamental problem, as argued in \cite{BeltranJimenez:2017doy}: the microscopic description of matter and radiation is provided by quantum fields, the expectation values of which are smooth. On the other hand, large gradients can potentially produce large effects---such effects can be exploited to place potentially strong constraints on model parameters and be useful for determining the phenomenological viability of the theory.

In this article, we develop the formalism of junction conditions and singular hypersurfaces in generalized coupling theories. In doing so, we establish a correspondence between the Einstein and Jordan frame expressions for the junction conditions and argue that the junction conditions do not lead to any obvious pathology so long as the derivatives are well behaved on each side of the junction surface. We then explore the problem in further detail in the MEMe model, examining, in particular, fluid profiles containing sharp gradients (as one might expect at the boundaries of horizonless ultracompact objects) in spherical symmetry. Both static and dynamical situations are examined; in the static case, we numerically solve the Tolman-Oppenheimer-Volkoff equation, studying the behavior of Jordan frame geodesics, and in the dynamical case we investigate pressureless collapse.

The rest of the present paper is organized as follows. In Sec \ref{sec:GC} we review and generalize the construction of generalized coupling theories. The formalism of singular hypersurfaces and junction conditions in generalized coupling theories is developed in Sec. \ref{sec:JSH}. In Sec. \ref{sec:MEMe}, we introduce the MEMe model and discuss the junction conditions. The problem of sharp gradients is explored in Sec. \ref{sec:SGrad}, and we summarize and discuss our results in Sec. \ref{sec:Conc}.

We employ the usual $\left(-,+,+,+\right)$ signature throughout. Subscripts and superscripts displayed as lowercase Greek letters denote coordinate basis indices, and those displayed as lowercase italic Latin denote field indices. We employ the convention that indices are raised and lowered exclusively with the Einstein frame metric $g_{\mu \nu}$. Gothic letters and hats indicate quantities defined with respect to the Jordan frame (for instance the metric $\mathfrak{g}_{\mu \nu}$ and its derivative $\hat{\nabla}$). Barred quantities indicate matrix inverses or in the case of Gothic letters, quantities with indices raised with the inverse Jordan frame metric $\bar{\mathfrak{g}}^{\mu \nu}$.

%=======================================================================

%-----------------------------------------------------------------------
%-----------------------------------
%-----------------
%--------
%---
%-
%
%
%-
%---
%--------
%-----------------
%-----------------------------------
%-----------------------------------------------------------------------

%=======================================================================
%-----------------------------------------------------------------------
%
%		GENERALIZED COUPLING THEORIES
%
%-----------------------------------------------------------------------
%=======================================================================
\section{Generalized coupling theories}\label{sec:GC}

In this section we briefly review generalized coupling theories and generalize slightly the action and formalism presented in \cite{Feng:2019dwu,Feng:2020lhp}.

\subsection{Action and field equations}
The action for a generalized coupling theory has the form
\begin{equation} \label{GC:Action}
 S_{\rm tot} = 
    S_{\rm g}[g^{\cdot \cdot}] 
    + 
    S_{\rm m}[\varphi,\bar{\mathfrak{g}}^{\cdot \cdot}]
    + 
    S_{\rm A}[g^{\cdot \cdot},A{_\cdot}{^\cdot}], 
\end{equation}
where $S_{\rm g}$ is the gravitational action, which is a functional of the inverse ``Einstein frame'' metric $g^{\mu \nu}$, and $S_{\rm m}$ is the matter action, which may be written as a functional of the matter fields $\varphi=\{\varphi_a\}$ and the inverse ``Jordan frame'' metric $\bar{\mathfrak{g}}^{\mu \nu}$. We assume that $\bar{\mathfrak{g}}^{\mu \nu}$ itself is a functional of $g^{\mu \nu}$ and a rank-2 tensor (termed the coupling tensor) $A{_\alpha}{^\beta}$,
\begin{equation} \label{GC:JmetricFunctional}
 \bar{\mathfrak{g}}^{\mu \nu} = \bar{\mathfrak{g}}^{\mu\nu}[{g}^{\cdot\cdot},{A}{_\cdot}{^\cdot}]
\end{equation}
The variation takes the form
\begin{equation} \label{GC:JmetricFunctionalDef}
   \resizebox{\columnwidth}{!}{$
   \delta \bar{\mathfrak{g}}^{\mu \nu} = \int d^4 y 
   \left[ 
   \mathcal{J}{^{\mu \nu}}{_{\alpha \beta}}(x,y)\delta {g}^{\alpha \beta}(y)
   +
   \mathcal{B}{^{\mu \nu \alpha}}{_{\beta}}(x,y)\delta {A}{_\alpha}{^\beta}(y)
   \right]
   $}
\end{equation}
where the tensor densities are the functional derivatives of ${g}^{\mu \nu}$ and ${A}{_\alpha}{^\beta}$
\begin{equation} \label{GC:JmetricFunctionalDerivDefs2}
   \begin{aligned}
   \mathcal{J}{^{\mu \nu}}{_{\alpha \beta}}(x,y) &:= \left.\frac{\delta \bar{\mathfrak{g}}^{\mu \nu}(x)}{\delta {g}^{\alpha \beta}(y)}\right|_{{A}{_\alpha}{^\beta}}, \\
   \mathcal{B}{^{\mu \nu \alpha}}{_{\beta}}(x,y) &:= \left.\frac{\delta \bar{\mathfrak{g}}^{\mu \nu}(x)}{\delta {A}{_\alpha}{^\beta}(y)} \right|_{{g}^{\alpha \beta}}.
   \end{aligned}
\end{equation}
For later convenience, we introduce the following notation for the variations of $S_{\rm g}$, $S_{\rm m}$, and $S_{\rm A}$ (neglecting boundary terms)
    \begin{align}
    \delta S_{\rm g} 
        & = - \frac{1}{2} \int d^4 x \sqrt{|g|} \, \mathcal{G}{_{\mu \nu}} \, \delta g^{\mu \nu}, \nonumber \\
    \delta S_{\rm m} 
        & = \frac{1}{2} \int d^4 x \sqrt{|{\mathfrak{g}}|} \left( \mathfrak{T}{_{\mu \nu}} \, \delta \bar{\mathfrak{g}}^{\mu \nu} + \mathcal{E}^a \delta \varphi_a + C{^\alpha}{_\beta} \, \delta A{_\alpha}{^\beta} \right), \nonumber \\
    \delta S_{\rm A} 
        & = \frac{1}{2} \int d^4 x \sqrt{|g|} \,  \left(
        t{_{\mu \nu}} \, \delta g^{\mu \nu} + \tilde{C}{^\alpha}{_\beta} \, \delta A{_\alpha}{^\beta} \right). \label{GC:VarActions}
\end{align}

\noindent The following quantities are defined
   \begin{align}
   & \mathcal{G}{_{\mu \nu}} := - \frac{2}{\sqrt{|g|}} \frac{\delta S_{\rm g}}{\delta g^{\mu \nu}} , \qquad \,
   \mathfrak{T}{_{\mu \nu}} := \left. \frac{2}{\sqrt{|\mathfrak{g}|}} \frac{\delta S_{\rm m}}{\delta \bar{\mathfrak{g}}^{\mu \nu}} \right|_{\varphi} , \nonumber\\
   & 
   \mathcal{E}^a := \left. \frac{2}{\sqrt{|\mathfrak{g}|}} \frac{\delta S_{\rm m}}{\delta \varphi_a} \right|_{\bar{\mathfrak{g}}^{\mu \nu}} , \qquad \quad \>
   t{_{\mu \nu}} := \left. \frac{2}{\sqrt{|g|}} \frac{\delta S_{\rm A}}{\delta g^{\mu \nu}} \right|_{\varphi} , \label{GC:JmetricFunctionalDerivDefs} \\
   & 
   {C}{^\alpha}{_\beta} := \left. \frac{2}{\sqrt{|g|}} \frac{\delta S_{\rm m}}{\delta A{_\alpha}{^\beta}} \right|_{\bar{\mathfrak{g}}^{\mu \nu}} , \qquad
   \tilde{C}{^\alpha}{_\beta} := \left. \frac{2}{\sqrt{|g|}} \frac{\delta S_{\rm A}}{\delta A{_\alpha}{^\beta}} \right|_{\bar{\mathfrak{g}}^{\mu \nu}} . \nonumber
   \end{align}
\noindent Upon substituting \eqref{GC:JmetricFunctionalDef} into \eqref{GC:VarActions}, the variation of $S_{\rm M}:=S_{\rm m}+S_{\rm A}$ takes the form
\begin{align} 
    &\resizebox{\columnwidth}{!}{$
    \delta S_{\rm M} 
    = \frac{1}{2} \int d^4 x \sqrt{|g|} \left( {T}{_{\mu \nu}} \delta {g}^{\mu \nu} + \mathcal{E}{^\alpha}{_\beta} \delta A{_\alpha}{^\beta} + \mathcal{E}^a \delta \varphi_a \sqrt{\mathfrak{g}/g} \right), 
     $}\label{GC:VarActionexplicit}
\end{align}
where
\begin{equation} \label{GC:VarActionexplicitTE}
   \begin{aligned}
   T_{\mu \nu} 
      & = t_{\mu \nu} + \frac{1}{\sqrt{|g|}} \int d^4 y \sqrt{|\mathfrak{g}(y)|} 
          \left[ 
            \mathfrak{T}_{\sigma \tau}(y) \, \mathcal{J}{^{\sigma \tau}}{_{\mu \nu}}(y,x) 
          \right],\\
\mathcal{E}{^\alpha}{_\beta}
      & = \tilde{C}{^\alpha}{_\beta} + \frac{1}{\sqrt{|g|}} \int d^4 y \sqrt{|\mathfrak{g}(y)|} 
          \left[ 
            \mathfrak{T}_{\sigma \tau}(y) \, \mathcal{B}{^{\sigma \tau \alpha}}{_{\beta}}(y,x)
          \right].
   \end{aligned}
\end{equation}
The complete set of equations has the form
\begin{equation} \label{GC:equationofmotion}
   \begin{aligned}
   \mathcal{G}{_{\mu \nu}} &= T{_{\mu \nu}}, \qquad
   \mathcal{E}{^\alpha}{_\beta} &= 0 , \qquad
   \mathcal{E}^a &= 0 .
   \end{aligned}
\end{equation}

\noindent If no derivatives of $A{_\mu}{^\alpha}$ are present in the action, then the field equation $\mathcal{E}{^\alpha}{_\beta} = 0$ is an algebraic equation relating $A{_\mu}{^\alpha}$ and the energy-momentum tensor $\mathfrak{T}{_{\mu \nu}}$.

%
%=======================================================================

%-----------------------------------------------------------------------
%		ENERGY-MOMENTUM CONSERVATION
%-----------------------------------------------------------------------
\subsection{Energy-momentum conservation}
Here, we derive the conservation of the energy-momentum tensor in both the Einstein and Jordan frames.\footnote{A standard derivation for ordinary general relativity may be found in Appendix E of \cite{Wald:1984rg}.}  We now assume that the actions $S_{\rm g}$ and $S_{\rm M}=S_{\rm m}+S_{\rm_A}$ are independently invariant under the infinitesimal diffeomorphism
\begin{equation} \label{GC:diffeomorphism}
   x^\mu \rightarrow x^\mu + \xi^\mu ,
\end{equation}

\noindent where the generating vector field $\xi$ is assumed to have compact support (in particular, we require it to vanish at the boundary of the integration domain). The infinitesimal changes in the fields may be written in terms of the Lie derivatives ($\delta \rightarrow \pounds_\xi$)
\begin{align}
   & \delta \varphi^a = \pounds_\xi \varphi^a , \nonumber \\
   & \delta g^{\mu \nu} = - 2 \nabla^{(\mu} \xi^{\nu)} , \label{GC:diffeovariations} \\
   & \delta A{_\alpha}{^\beta} = \xi^\sigma \nabla_\sigma A{_\alpha}{^\beta} - A{_\alpha}{^\sigma} \nabla_\sigma \xi^\beta + A{_\sigma}{^\beta} \nabla_\alpha \xi^\sigma . \nonumber
\end{align}

\noindent The change in the gravitational action $S_{\rm g}[g^{\cdot \cdot}]$ due to an arbitrary $\xi^\mu$ of compact support has the form
\begin{equation} \label{GC:VarActionGexplicitdiffeo}
   \begin{aligned}
   \delta S_{\rm g} 
      & = \int d^4 x \sqrt{|g|} \, \xi^{\nu} \nabla^{\mu} \mathcal{G}{_{\mu \nu}},
   \end{aligned}
\end{equation} 

\noindent and requiring $\delta S_{\rm g}=0$ we recover the result $\nabla^\mu \mathcal{G}{_{\mu \lambda}} = 0$. The change in the action $S_{\rm M}:=S_{\rm m} + S_{\rm_A}$ due to the infinitesimal diffeomorphism has the form (after performing an integration by parts)
\begin{equation} \label{GC:VarActionMexplicitdiffeo}
   \begin{aligned}
   \delta S_{\rm M}
        & = \frac{1}{2} \int d^4 x \sqrt{|g|} \, \xi^\lambda \biggl[ 2 \nabla^\mu {T}{_{\mu \lambda}} + \mathcal{E}{^\alpha}{_\beta} \nabla_\lambda A{_\alpha}{^\beta} \\
        & \quad + \nabla_\sigma (\mathcal{E}{^\alpha}{_\lambda} A{_\alpha}{^\sigma}) - \nabla_\alpha ( \mathcal{E}{^\alpha}{_\beta} A{_\lambda}{^\beta}) -\mathcal{D}^*_{a \lambda} (\varphi,\nabla) \, \underline{\mathcal{E}}^a \biggr],
   \end{aligned}
\end{equation}

\noindent where we have defined the quantity
\begin{equation}\label{GCA-EvarphiRescaled}
   \underline{\mathcal{E}}^a:=\sqrt{\mathfrak{g}/g} \, \mathcal{E}^a.
\end{equation}

\noindent Here, we define the linear differential operator $\mathcal{D}^*_{a \lambda} (\varphi,\nabla)$ for some field $\varphi_a$ and connection $\nabla$ by the following expression
\begin{equation}\label{GCA-LinearDifferentialOperator}
   \pounds_\xi \varphi_a = \mathcal{D}_{a \lambda} (\varphi,\nabla) \xi^\lambda ,
\end{equation}

\noindent and its Hermitian conjugate (which is also a linear differential operator)
\begin{equation}\label{GCA-LinearDifferentialOperatorHC}
   \int d^4 x \sqrt{|g|} \, f^a \mathcal{D}_{a \lambda} (\varphi,\nabla)\xi^\lambda
   =
   \int d^4 x \sqrt{|g|} \, \xi^\lambda \mathcal{D}^*_{a \lambda} (\varphi,\nabla) f^a ,
\end{equation}

\noindent with $f^a$ being smooth test functions of compact support. For a scalar field $\phi$, one has 
\begin{equation}
\begin{split}
    \mathcal{D}_\lambda (\cdot)^\lambda &=  (\cdot)^\lambda \partial_\lambda \phi,\\
    \mathcal{D}^*_{\lambda}(\cdot) &= (\cdot)\partial_{\lambda}\phi,
\end{split}    
\end{equation}
and for a one-form $\omega_\mu$, one has 
\begin{equation}
\begin{split}
    \mathcal{D}_{\mu \lambda}(\cdot)^\lambda &=  (\cdot)^\lambda \nabla_\lambda \omega_\mu + \omega_{\lambda} \nabla_\mu (\cdot)^\lambda,\\
    \mathcal{D}^*_{\mu \lambda} (\cdot)^\mu &= (\cdot)^\mu \nabla_\lambda \omega_\mu - (\cdot)^\mu \nabla_\mu \omega_{\lambda} - \omega_{\lambda} \nabla_\mu (\cdot)^\mu .
\end{split}    
\end{equation}
We note that since $\mathcal{D}^*_{\mu \lambda}$ is a linear operator, $\mathcal{D}^*_{\mu \lambda}f^\mu$ vanishes if the test function $f^\mu$ vanishes everywhere. Requiring that $\delta S_m = 0$ for an arbitrary $\xi^\mu$ of compact support, we find
\begin{equation}\label{GCA-EMTDivergence}
    \begin{aligned}
    2 \nabla^\mu {T}{_{\mu \lambda}} = \, &
        - \mathcal{E}{^\alpha}{_\beta} \nabla_\lambda A{_\alpha}{^\beta} 
        - \nabla_\sigma (\mathcal{E}{^\alpha}{_\lambda} A{_\alpha}{^\sigma}) \\
        &
        + \nabla_\alpha ( \mathcal{E}{^\alpha}{_\beta} A{_\lambda}{^\beta}) + \mathcal{D}^*_{a \lambda} (\varphi,\nabla) \, \underline{\mathcal{E}}^a . 
   \end{aligned}
\end{equation}

\noindent On solutions of the field equations $\mathcal{E}{^\alpha}{_\beta}=0$ and $\underline{\mathcal{E}}^a=0$, we recover the conservation of the Einstein frame energy-momentum tensor $\nabla^\mu {T}{_{\mu \lambda}} = 0$. 

So far, we have obtained these results for Einstein frame variables. However, matter is assumed to be minimally coupled to the Jordan frame metric $\mathfrak{g}_{\mu \nu}$ and its inverse $\bar{\mathfrak{g}}^{\mu \nu}$, so one might expect similar results to hold with respect to Jordan frame variables. The infinitesimal change with respect to the Jordan frame variables may be written as
\begin{equation} \label{GC:diffeovariationsJordan}
   \begin{aligned}
    & \delta \bar{\mathfrak{g}}_{\mu \nu} = - 2 \hat{\nabla}_{(\mu} \xi_{\nu)} , \\
    & \delta A{_\alpha}{^\beta} = \xi^\sigma \hat{\nabla}_\sigma A{_\alpha}{^\beta} - A{_\alpha}{^\sigma} \hat{\nabla}_\sigma \xi^\beta + A{_\sigma}{^\beta} \hat{\nabla}_\alpha \xi^\sigma ,
   \end{aligned}
\end{equation}

\noindent where the Jordan frame Levi-Civita connection $\hat{\nabla}$ is defined by $\hat{\nabla}_\sigma \mathfrak{g}_{\mu \nu}=0$. After an integration by parts, the variation of the matter action $S_{\rm m}$ has the form (where $\bar{\mathfrak{T}}{^{\alpha \beta}} = \bar{\mathfrak{g}}{^{\mu \alpha}} \bar{\mathfrak{g}}{^{\nu \beta}} {\mathfrak{T}}{_{\mu \nu}}$):
\begin{equation} \label{GC:MatterActionVariationDiffeo}
    \begin{aligned}
    \delta S_{\rm m} 
        & = \frac{1}{2} \int d^4 x \sqrt{|\mathfrak{g}|} \, \xi^{\lambda} \left( 2 \mathfrak{g}_{\lambda \beta} \hat{\nabla}_{\alpha}\bar{\mathfrak{T}}{^{\alpha \beta}} -\mathcal{D}^*_{a \lambda} (\varphi,\hat\nabla) \, \mathcal{E}^a \right),
   \end{aligned}
\end{equation}

\noindent and upon requiring $\delta S_{\rm m} = 0$ for an arbitrary $\xi^\mu$ of compact support, we obtain the expression
\begin{equation} \label{GC:JordanFrameConservationLaw}
\hat{\nabla}_{\alpha}\bar{\mathfrak{T}}{^{\alpha \beta}} = \tfrac{1}{2} \bar{\mathfrak{g}}^{\lambda \beta} \mathcal{D}^*_{a \lambda} (\varphi,\hat\nabla) \, \mathcal{E}^a .
\end{equation}

\noindent On solutions to the field equations $\mathcal{E}^a=0$, one recovers the conservation law in the Jordan frame, $\hat{\nabla}_{\alpha}\bar{\mathfrak{T}}{^{\alpha \beta}} = 0$.

%=======================================================================

%-----------------------------------------------------------------------
%		INTEGRATING OUT NONDYNAMICAL COUPLING TENSORS
%-----------------------------------------------------------------------
\subsection{Integrating out nondynamical coupling tensors}
If the coupling tensor $A{_\alpha}{^\beta}$ is nondynamical (in the sense that it satisfies algebraic, rather than differential equations of motion), one can integrate it out of the equations of motion. In particular, we consider the case where $\mathcal{E}{^\alpha}{_\beta} = 0$ is an algebraic equation for $A{_\alpha}{^\beta}$ with a unique solution of the form
\begin{equation} \label{GC:Asolabstract}
 A{_\alpha}{^\beta} = \underline{A}{_\alpha}{^\beta} \left[\varphi,g^{\cdot \cdot}\right] .
\end{equation}

\noindent Since the equation of motion is algebraic, one can substitute the above solution in the action $S=S[g^{\cdot\cdot},A{_\cdot}{^\cdot},\varphi]$ to obtain
\begin{equation} \label{GC:IntegratedOutAction}
\underline{S}[g^{\cdot\cdot},\varphi] = S[g^{\cdot\cdot},\underline{A}{_\cdot}{^\cdot}[\varphi,g^{\cdot\cdot}],\varphi],
\end{equation}

\noindent Upon performing the variation of the above action $\underline{S}[g^{\cdot\cdot},\varphi]$, one finds that the equations of motion have the form
\begin{equation} \label{GC:equationofmotionIntegratedOut}
   \begin{aligned}
   \left. \mathcal{G}{_{\mu \nu}} = T{_{\mu \nu}} \right|_{A{_\alpha}{^\beta} = \underline{A}{_\alpha}{^\beta}}, \qquad
   \left. \mathcal{E}^a = 0 \right|_{A{_\alpha}{^\beta} = \underline{A}{_\alpha}{^\beta}}.
   \end{aligned}
\end{equation}
where we have used the fact that $\left.\mathcal{E}^{\gamma}_{\ \delta}\right|_{A_{\alpha}^{\ \beta}=\underline{A}_{\alpha}^{\ \beta}}=0$.

The action \eqref{GC:IntegratedOutAction} and the corresponding equations of motion \eqref{GC:equationofmotionIntegratedOut} do not contain $A_{\alpha}^{\beta}$ and are equivalent to those in the original theory. To be more precise, the set of all solutions of \eqref{GC:equationofmotionIntegratedOut} supplemented by \eqref{GC:Asolabstract} is the same as the set of all solutions of \eqref{GC:equationofmotion}. 

%=======================================================================

%-----------------------------------------------------------------------
%		RICCI CURVATURE IN THE JORDAN FRAME
%-----------------------------------------------------------------------
\subsection{Ricci curvature in the Jordan frame}
We now consider a case in which $\mathcal{G}_{\mu \nu}$ can be written in terms of the Ricci curvature tensor, and in which Einstein and Jordan frame metrics are related by the following expression
\begin{equation}\label{GCA-EJInvMetricRelation}
   {g}^{\mu \nu} = \Psi \,  A{_\alpha}{^\mu} \, A{_\beta}{^\nu} \, \bar{\mathfrak{g}}^{\alpha \beta} ,
\end{equation}
\begin{equation}\label{GCA-EJMetricRelation}
   {g}_{\mu \nu} = \Psi^{-1} \, \bar{A}{^\alpha}{_\mu} \, \bar{A}{^\beta}{_\nu} \, \mathfrak{g}_{\alpha \beta} ,
\end{equation}

\noindent where $\Psi=\Psi(A{_\cdot}{^\cdot})$ is a scalar made of $A{_\alpha}{^\mu}$ and the nondegenerate tensors $A{_\alpha}{^\mu}$ and $\bar{A}{^\alpha}{_\mu}$ are inverses of each other. Here, we follow the convention that indices are raised and lowered exclusively with the Einstein frame metric.

A relationship between the connections may be established by the following relationship between the connection coefficients
\begin{equation}\label{GCA-EJConnectionRelation}
\begin{aligned}
   \Gamma{^\sigma}_{\mu \nu} 
   & = \hat{\Gamma}{^\sigma}_{\mu \nu} + W^{\sigma}{}_{\mu \nu} ,
\end{aligned}
\end{equation}

\noindent where $\hat{\Gamma}{^\sigma}_{\mu \nu}$ are the Levi-Civita connection coefficients associated with the connection $\hat{\nabla}$, and the contorsion tensor $W^{\sigma}{}_{\mu \nu}$ takes the form
\begin{equation}\label{GCA-W}
   \begin{aligned}
      W^{\sigma}{}_{\mu \nu}
      & = 
      \frac{1}{2} g^{\sigma \tau} \left(\hat{\nabla}_\mu g_{\nu \tau} + \hat{\nabla}_\nu g_{\mu \tau} - \hat{\nabla}_\tau g_{\mu \nu} \right) \\
      & = 
      Q{^\sigma}{_{\mu \nu}}{^{\tau \beta}}_\alpha \, \hat{\nabla}_\tau \bar{A}{^\alpha}{_\beta}
      + \mathcal{Q}{^\sigma}{_{\mu \nu}}{^\tau} \, \hat{\nabla}_\tau \Psi
      ,
   \end{aligned}
   \end{equation}

\noindent where the coefficients are given by
\begin{equation}\label{GCA-EJConnectionRelationQs}
   \begin{aligned}
      Q{^\sigma}{_{\mu \nu}}{^{\tau \beta}}_\alpha 
      & = 
      A_{\alpha}{}^{\sigma} \delta _{\mu}{}^{(\tau} \delta _{\nu}{}^{\beta)} 
      +
      2A_{\alpha}{}^{\rho} g_{\rho (\mu}  \delta_{\nu)}{}^{[\tau} g^{\beta] \sigma},\\
      \mathcal{Q}{^\sigma}{_{\mu \nu}}{^\tau}
      & =
      \frac{1}{2 \Psi} (
        \delta _{\alpha}{}^{\tau} g_{\mu \nu} 
        - 2 \delta _{[\mu}{}^{\tau} g_{\nu] \alpha}
      ) g^{\sigma \alpha} .
   \end{aligned}
\end{equation}

\noindent The Ricci tensor takes the form
\begin{equation}\label{GCA-RicciTensorW}
   \begin{aligned}
      {R}_{\mu \nu}
      = \, &
      \hat{R}_{\mu \nu}
      + 
      \hat{\nabla }_{\sigma} W^{\sigma}{}_{\mu \nu} - \hat{\nabla }_{\mu}W^{\sigma}{}_{\sigma \nu} \\
      &
      + 
      W^{\sigma}{}_{\mu \nu} W^{\tau}{}_{\tau \sigma} - W^{\sigma}{}_{\mu}{}^{\tau} W_{\tau \sigma \nu} .
   \end{aligned}
\end{equation}

\noindent Explicitly, one may rewrite the Ricci tensor as
\begin{equation}\label{GCA-RicciTensorW2}
   \begin{aligned}
      {R}_{\mu \nu}
      =\,&
      \hat{R}_{\mu \nu}
      + 
      \left( \mathcal{Q}{^\sigma}{_{\mu \nu}}{^\tau} - \delta {_\mu}{^\sigma} \mathcal{Q}{^\rho}{_{\rho \nu}}{^\tau} \right) \hat{\nabla }{_\sigma}\hat{\nabla }{_\tau} \Psi  \\
      &
      + 
      \left( Q^{\sigma}{}{_{\mu \nu}}{}^{\tau \delta}{}{_\gamma} - \delta {_\mu}{}^{\sigma} Q^{\rho}{}{_{\rho \nu}}{}^{\tau \delta}{}{_\gamma} \right) \hat{\nabla }{_\sigma}\hat{\nabla }{_\tau}\bar{A}^{\gamma}{}{_\delta} \\
      &
      + 
      {M} {_{\mu \nu}}{}^{\delta \zeta}{}{_\epsilon}{^{\lambda \chi}}_{\kappa} \> \hat{\nabla}_\lambda \bar{A}{^\kappa}{_\chi} \hat{\nabla }{_\delta}\bar{A}^{\epsilon}{}{_\zeta} \\
      &
      + 
      {N} {_{\mu \nu}}{^{ \sigma \lambda \chi}}_{\kappa} \, \hat{\nabla}_\lambda \bar{A}{^\kappa}{_\chi} \hat{\nabla }{_\sigma}\Psi  
      + 
      {P} {_{\mu \nu}}{^{ \sigma \tau}} \hat{\nabla}_\tau \Psi \hat{\nabla }{_\sigma}\Psi
      ,
   \end{aligned}
\end{equation}

\noindent with the following expressions for the coefficients (presented here for completeness)
\begin{equation}\label{GCA-DQCs}
   \begin{aligned}
      M {_{\mu \nu}}{}^{\delta \zeta}{}{_\epsilon}{^{\lambda \chi}}_{\kappa}
      = & \,
         A_{\epsilon \rho} A_{\kappa \tau}
         \biggl[
            2 \delta _{\mu}{}^{(\delta} \delta _{\nu}{}^{\zeta)}
            g^{\lambda (\chi} g^{\tau) \rho}
            + g^{\rho \tau}
            \delta_{\mu}{}^{\lambda} \delta_{\nu}{}^{\delta} g^{\zeta\chi}\\
         & 
         -
            2 \delta _{\mu}{}^{\lambda} \delta _{\nu}{}^{\delta} 
            g^{\rho (\chi} g^{\tau) \zeta} + 4 \delta_{\sigma}{^\tau} \delta_{(\mu}{^\rho} 
            \delta _{\nu)}{}^{[\delta} g^{\zeta] (\sigma} g^{\chi) \lambda}
          \\
         &
         + g^{\rho \tau}
         \left(
         2\delta_{\mu}{}^{(\chi} \delta_{\nu}{}^{[\zeta)} g^{\delta]\lambda}
         -
         \delta_{\mu}{}^{\delta} \delta_{\nu}{}^{(\zeta} g^{\chi)\lambda}
         \right)
        \biggr]\\
        &
        + 
        2 Q^{\alpha}{}_{\mu [\nu|}{}^{ \lambda \chi}{}_{\kappa} Q^{\beta}{}_{|\beta] \alpha}{}^{\delta \zeta}{}_{\epsilon} ,\\
      N {_{\mu \nu}}{^{ \sigma \lambda \chi}}_{\kappa}
      = &\frac{1}{\Psi}
      \biggl\{
            A_{\kappa}{}^{\alpha} \delta_{(\mu}{}^{\chi} g_{\nu) \alpha} g^{\lambda \sigma} 
            - 
            g_{\mu \nu} 
            A_{\kappa}{}^{(\sigma} g^{\lambda) \chi}
      \biggr\} \\
      &
      + 2 \mathcal{Q}^{\alpha}{}_{\beta [\alpha}{}^{\sigma} Q^{\beta}{}_{\nu] \mu}{}^{\lambda \chi}{}_{\kappa}
      + 2 \mathcal{Q}^{\alpha }{}_{\mu [\nu}{}^{\sigma} Q^{\beta }{}_{\beta] \alpha }{}^{\lambda \chi}{}_{\kappa} ,\\
      P{_{\mu \nu}}{^{ \sigma \tau}}
      = &\frac{1}{2 \, \Psi^2} 
      \biggl\{
        2(\delta _{\mu}{}^{[\sigma} \delta _{\nu}{}^{\tau]} - \delta _{\mu}{}^{\tau} \delta _{\nu}{}^{\sigma}) - g_{\mu \nu} g^{\tau \sigma}
      \biggr\} \\
      &
      + 2 \mathcal{Q}^{\alpha}{}_{\mu [\nu}{}^{\tau} \mathcal{Q}^{\beta}{}_{\beta] \alpha}{}^{\sigma}. 
\end{aligned}
\end{equation}

\noindent If $\mathcal{G}_{\mu \nu}$ depends exclusively on the Ricci tensor constructed from the Einstein frame metric $g_{\mu \nu}$, one can in principle use the expression in Eq. \eqref{GCA-RicciTensorW2} to rewrite the gravitational field equations in terms of the metric $\mathfrak{g}_{\mu \nu}$, the coupling tensor $A{_\alpha}{^\mu}$, and the factor $\Psi$. Of course, we must point out that to simplify the coefficients, we have lowered the second index of the coupling tensors $A_{\mu}{}^{\alpha}$ in the expression for $M {_{\mu \nu}}{}^{\delta \zeta}{}{_\epsilon}{^{\lambda \chi}}_{\kappa}$; to rewrite the Ricci tensor in terms of Jordan frame quantities, one should take care to restore the indices to their proper positions and replace $g_{\mu \nu}$ and $g^{\mu \nu}$ according to Eqs. \eqref{GCA-EJInvMetricRelation} and \eqref{GCA-EJMetricRelation}.

This expression for the Ricci curvature illustrates the concerns brought up in \cite{Pani:2012qd,Pani:2013qfa}. Sharp gradients in the energy-momentum tensor $\mathfrak{T}{_{\mu \nu}}$ will generally induce sharp gradients in the coupling tensor $A{_\mu}{^\alpha}$ and its inverse $\bar{A}{^\alpha}_\mu$. Discontinuities in the coupling tensor $A{_\mu}{^\alpha}$ will in general lead to curvature singularities in either the Jordan or Einstein frame Ricci tensor. In the Einstein frame, the field equations \eqref{GC:equationofmotion} contain no derivatives of $\mathfrak{T}{_{\mu \nu}}$, so even if there are discontinuities in $\mathfrak{T}{_{\mu \nu}}$, the Einstein frame Ricci tensor does not necessarily become singular at the surfaces of discontinuity; in this way, generalized coupling theories evade the pathologies indicated in \cite{Pani:2012qd,Pani:2013qfa}. On the other hand, if one rewrites the Einstein frame equations \eqref{GC:equationofmotion} in terms of Jordan frame variables, the equations contain singular terms; the Jordan frame Ricci curvature $\hat{R}_{\mu \nu}$, for instance, will contain singularities in the presence of discontinuities in $\mathfrak{T}{_{\mu \nu}}$, and it is not apparent that the Einstein frame metric $g_{\mu \nu}$, as constructed from $\mathfrak{g}_{\mu \nu}$ and $A{_\mu}{^\alpha}$, yields well-behaved curvature tensors. For this reason, it is convenient to regard the primary gravitational degrees of freedom as being contained in the Einstein frame metric $g_{\mu \nu}$.

%=======================================================================

%-----------------------------------------------------------------------
%-----------------------------------
%-----------------
%--------
%---
%-
%
%
%-
%---
%--------
%-----------------
%-----------------------------------
%-----------------------------------------------------------------------

%=======================================================================
%-----------------------------------------------------------------------
%
%		SINGULAR HYPERSURFACES
%
%-----------------------------------------------------------------------
%=======================================================================
\section{Singular hypersurfaces and Junction conditions}\label{sec:JSH}

In this section, we develop the formalism of singular hypersurfaces and junction conditions for generalized coupling theories. However, before jumping into the formalism straight away, it is appropriate to first establish a physical understanding of singular hypersurfaces and discuss our strategy for developing the formalism. To simplify the discussion and the formalism, we restrict our considerations to a class of generalized coupling theories which admits an Einstein frame, assuming that $S_{\rm g}[g^{\cdot \cdot}]$ is the Einstein-Hilbert action; in this case, the theory is equivalent to general relativity with a modified matter action. We also suppose that $\mathcal{E}^{\alpha}_{\ \beta}=0$ is an algebraic equation for $A_{\alpha}^{\ \beta}$ and admits a unique solution of the form (\ref{GC:Asolabstract}).

In general relativity, singular hypersurfaces and junction conditions are described by the Israel formalism \cite{Israel:1966rt,Poisson:2009pwt}, so we first consider a singular hypersurface in the Einstein frame (which is also appropriate in light of the discussion at the end of the preceding section). Needless to say, the hypersurface is not literally singular. Instead, we consider a thin layer with a thickness that is sufficiently larger than the cutoff length scale of the theory and that is sufficiently smaller than length scales of physical interest. For any such choices of thickness, the system can be described by general relativity with a modified matter action. The Israel junction condition in general relativity then implies that the induced metric computed from the Einstein-frame metric $g_{\mu\nu}$ should be continuous across the hypersurface. On the other hand, $A_{\alpha}^{\ \beta}$ given by (\ref{GC:Asolabstract}) does not have to be continuous, meaning that the induced metric in the Jordan frame can be discontinuous. This is because the Jordan-frame metric is an effective metric that is dressed by the stress-energy tensor of matter fields and the stress-energy tensor can be discontinuous across the singular hypersurface.\footnote{For example, for a canonical scalar field, the stress-energy tensor contains first derivatives of the scalar field and can be discontinuous if there is a non-trivial potential for the scalar field localized on the hypersurface.} Despite the discontinuity of the Jordan-frame metric, the matter dynamics can be solved consistently since the matter equations of motion can be rewritten in terms of the continuous Einstein-frame metric [see (\ref{GC:equationofmotionIntegratedOut})]. With this perspective in mind, it is appropriate to first develop the formalism for singular hypersurfaces and junctions in the Einstein frame.

%=======================================================================

%-----------------------------------------------------------------------
%		VARIATION IN THE EINSTEIN FRAME
%-----------------------------------------------------------------------
\subsection{Variation in the Einstein frame}
Upon splitting the manifold $U = U_{-} \cup U_{+}$ into two regions $U_{-}$ and $U_{+}$ separated by a codimension one junction surface $\Sigma$, we introduce in each region the metric $g^\pm_{\mu\nu}$ (and its inverse $g_\pm^{\mu\nu}$), the Ricci scalar $R_{\pm}$, the cosmological constant $\Lambda_{\pm}$ and the unit vector $n_{\pm}^{\mu}$ normal to $\Sigma$. We then obtain the following action,

\begin{equation} \label{GC:ActionGrav}
   \begin{aligned}
   S_{\rm g}[{g}^{\cdot \cdot}] = 
   \frac{1}{2 \kappa} 
   \biggl[ 
      &
      \int_{U_{-}} d^4 x \sqrt{|g^{-}|} \left( R_{-} - 2 \Lambda_{-} \right) \\
      &
      + 
      \int_{U_{+}} d^4 x \sqrt{|g^{+}|} \left( R_{+} - 2 \Lambda_{+} \right) \\
      &
      +
      \epsilon \int_{\Sigma} d^3y 
      \biggl\{2\left\llbracket\sqrt{|q|} \, K \right\rrbracket + \lambda_{\mu \nu}\left\llbracket q^{\mu \nu}\right\rrbracket \biggr\}
   \biggr] ,
   \end{aligned}
\end{equation}

\noindent where $\lambda_{\mu\nu}$ are Lagrange multipliers enforcing the continuity of the induced metric~\cite{Mukohyama:2001pb} and the double-struck brackets $\llbracket \, \rrbracket$ indicate differences in some quantity on each side of the junction surface $\Sigma$ (for instance $\llbracket Q \rrbracket:=Q_+-Q_-$). Here, $q^\pm_{\mu\nu} = g^\pm_{\mu\nu} - \varepsilon n^\pm_{\mu}n^\pm_{\nu}$ and $K_{\pm} =  \nabla^\pm_{\alpha} n_{\pm}^{\alpha}$ are respectively the induced metric and the trace of the extrinsic curvature on $\Sigma$ computed from each side, $\nabla_{\pm\alpha}$ is the covariant derivative compatible with $g_{\pm\mu\nu}$, and $\varepsilon=\pm 1$ is positive if the unit normal is spacelike and negative if the unit normal is timelike. 

The action $S_{\rm M}=S_{\rm m}+S_{\rm A}$ splits in the following manner,
\begin{equation} \label{GC:ActionMatt}
\begin{aligned}
   S_{\rm M}[\varphi,{g}^{\cdot \cdot},A{_\cdot}{^\cdot}]
   =
   &
   \int_{U_{-}} d^4 x \sqrt{|g^{-}|} \, \mathcal{L}_{\rm M-}[\varphi_-,{g}_-^{\cdot \cdot},A{^-_\cdot}{^\cdot}] \\
   &
   + 
   \int_{U_{+}} d^4 x \sqrt{|g^{+}|} \, \mathcal{L}_{\rm M+}[\varphi^+,{g}_+^{\cdot \cdot},A{^+_\cdot}{^\cdot}] \\
   &
   + 
   \int_{\Sigma} d^3y 
   \left(\hat{\mathcal{L}}_{\rm M,0}[\varphi^\pm,{q}_\pm^{\cdot \cdot},A{^\pm_\cdot}{^\cdot}]\right) .
\end{aligned}
\end{equation}

\noindent The total variation of the action $S:=S_{\rm g}+S_{\rm M}$, making use of Weiss variation methods [see appendix \eqref{GC:ActionGrav2bVar3} and \cite{Feng:2021lfa,Feng:2017ygy,Mukohyama:2001pb} for details], has the following form,
\begin{equation} \label{GC:ActionVar}
   \begin{aligned}
      \delta S
      =
      &
      \int_{U_{-}} d^4 x \sqrt{|g^{-}|} \, \biggl[ \mathcal{E}_-^a \delta \varphi^-_a + \mathcal{E}{_-^\mu}{_\nu} \delta A{^-_\mu}{^\nu} \\
      & \qquad \qquad \qquad \quad +
      \frac{1}{2\kappa} \left(G^-_{\mu \nu} + \Lambda_- g^-_{\mu \nu} - \kappa \, T^-_{\mu \nu} \right) \delta g_-^{\mu \nu}
       \biggr] \\
      &
      + 
      \int_{U_{+}} d^4 x \sqrt{|g^{+}|} \, \biggl[ \mathcal{E}_+^a \delta \varphi^+_a + \mathcal{E}{_+^\mu}{_\nu} \delta A{^+_\mu}{^\nu} \\
      & \qquad \qquad \qquad \quad +
      \frac{1}{2\kappa} \left(G^+_{\mu \nu} + \Lambda_+ g^+_{\mu \nu} - \kappa \, T^+_{\mu \nu} \right) \delta g_+^{\mu \nu}
       \biggr] \\
      &
      +
      \int_{\Sigma} d^3y
      \biggl\llbracket 
         \sqrt{|q|}\left\{ n^\mu \, P{^a}{_\mu} - s^a \right\} \delta \varphi_a  - \sqrt{|q|} \, s{^\mu}{_\nu} \, \delta A{_\mu}{^\nu} \\
         & 
         + 
         \kappa^{-1} \sqrt{|q|} \left\{G{^\mu}{_\nu} + \Lambda  \delta{^\mu}{_\nu} - \kappa \, \bar{T}{^\mu}{_\nu}\right\} n_\mu \, \delta Z^\nu \\
         &
         +
         \frac{\epsilon}{2 \kappa}\left\{  \sqrt{|q|} \left(K_{\mu \nu} - K q_{\mu \nu} - \kappa \epsilon  S_{\mu \nu} \right) + \lambda_{\mu \nu} \right\} \delta q^{\mu \nu} 
      \biggr\rrbracket\\
      &
      + \frac{\epsilon}{2 \kappa}\int_{\Sigma} d^3y \, \delta \lambda_{\mu \nu} \llbracket q^{\mu \nu}\rrbracket
      ,
   \end{aligned}
\end{equation}

\noindent where we have defined the functional derivatives $\sqrt{|g^\pm|} \mathcal{E}_\pm^a := {\delta S_{\rm M}}/{\delta \varphi^\pm_a}$ and $\sqrt{|g^\pm|} \mathcal{E}{_\pm^\mu}{_\nu}:={\delta S}/{\delta A{^\pm_\mu}{^\nu}}$, and the energy-momentum tensor has the usual definition $\sqrt{|g^\pm|} \, T^\pm_{\mu \nu} : = -2{\delta S_{\rm M}}/{\delta g_\pm^{\mu \nu}}$. The following surface functional derivatives have also been defined,
\begin{equation} \label{GC:ActionMattVarDefs}
   \begin{aligned}
        \sqrt{|q|} \, S^\pm_{\mu \nu} &: = -2\frac{\delta}{\delta q_\pm^{\mu \nu}} 
                        \int_{\Sigma} d^3y 
                        \left(\hat{\mathcal{L}}_{\rm M,0}[\varphi^\pm,{q}_\pm^{\cdot \cdot},A{^\pm_\cdot}{^\cdot}]\right) , \\
        \sqrt{|q|} \, s_\pm^a &: = \frac{\delta}{\delta \varphi^\pm_a} 
               \int_{\Sigma} d^3y 
               \left(\hat{\mathcal{L}}_{\rm M,0}[\varphi^\pm,{q}_\pm^{\cdot \cdot},A{^\pm_\cdot}{^\cdot}]\right) , \\
        \sqrt{|q|} \, s_\pm^{\mu}{_\nu} &: = \frac{\delta}{\delta A{^\pm_\mu}{^\nu}} 
                           \int_{\Sigma} d^3y 
                           \left(\hat{\mathcal{L}}_{\rm M,0}[\varphi^\pm,{q}_\pm^{\cdot \cdot},A{^\pm_\cdot}{^\cdot}]\right)
      .
   \end{aligned}
\end{equation}

\noindent The bulk field equations takes the form
\begin{equation} \label{GC:BulkFieldEqs}
   \begin{aligned}
      & \mathcal{E}_\pm^a = 0, \\
      & \mathcal{E}{_\pm^\mu}{_\nu} = 0, \\
      & G{^\pm_{\mu \nu}} + \Lambda_\pm \, g^\pm_{\mu \nu} = \kappa T{^\pm_{\mu \nu}}
      .
   \end{aligned}
\end{equation}

\noindent The variation with respect to the Lagrange multipliers $\lambda_{\mu \nu}$ yield the Israel junction condition
\begin{equation} \label{GC:BdyFieldEqsIsrael}
   \begin{aligned}
    &  \llbracket q^{\mu \nu} \rrbracket = 0.
   \end{aligned}
\end{equation}

\noindent If the Einstein frame boundary metric variations are assumed to be continuous, one has the boundary field equation (assuming $\llbracket\lambda_{\mu \nu}\rrbracket=0$),
\begin{equation} \label{GC:BdyFieldEqsMetric}
   \begin{aligned}
    & \kappa \llbracket \sqrt{|q|} \, S_{\mu \nu} \rrbracket = \varepsilon \left\llbracket \sqrt{|q|} \left(K_{\mu \nu} - K \, q_{\mu \nu}\right) \right\rrbracket ,
   \end{aligned}
\end{equation}

\noindent which may, with the help of \eqref{GC:BdyFieldEqsIsrael}, be rewritten in the trace-reversed form
\begin{equation} \label{GC:BdyFieldEqsMetricTR}
   \begin{aligned}
    & \llbracket K_{\mu \nu} \rrbracket = \frac{\kappa}{\varepsilon}  \left\llbracket S_{\mu \nu} - \frac{1}{2} S \, q_{\mu \nu} \right\rrbracket .
   \end{aligned}
\end{equation}

\noindent If the fields $\varphi_a$ and $A{_\mu}{^\nu}$ are not assumed to be continuous at the boundary surface, in particular, if $\delta \varphi^+$, $\delta \varphi^-$, $A{^+_\mu}{^\nu}$, and $A{^-_\mu}{^\nu}$ are all assumed to be independent of each other, one has the following field equations
\begin{equation} \label{GC:BdyFieldEqsDisc}
   \begin{aligned}
      & n_\pm^\mu \, P{_\pm^a}{_\mu} = s_\pm^a, \\
      & s{_\pm^\mu}{_\nu} = 0 .
   \end{aligned}
\end{equation}

\noindent On the other hand, if $\delta \varphi$ and $\delta A{_\mu}{^\nu}$ are assumed to be continuous, one has instead the surface field equations $\llbracket n_\mu \, P{_a}{^\mu} - s_a \rrbracket = 0$ and $\llbracket s{^\mu}{_\nu} \rrbracket = 0$.

%=======================================================================

%-----------------------------------------------------------------------
%		RELATING SURFACE EINSTEIN AND JORDAN FRAME QUANTITIES
%-----------------------------------------------------------------------
\subsection{Relating surface Einstein and Jordan frame quantities}
Since we have effectively two geometries on the spacetime manifold, one given by the metric $g_{\mu \nu}$ and the other by $\mathfrak{g}_{\mu \nu}$, it is perhaps appropriate to define the hypersurface in a manner that is independent of the metric geometry. In particular, one may imagine a manifold endowed with a coordinate chart $x^\mu$, then define a hypersurface as a level surface of some scalar function $\Phi(x)$ of the coordinates. The gradient is defined as follows:
\begin{equation}\label{GCA-DualNormal}
   \Phi_\mu := \nabla_\mu \Phi.
\end{equation}

\noindent The quantity $\Phi_\mu$ is defined purely in terms of the coordinates. Given a metric, one finds that $\Phi_\mu$ (assuming it is non-null) is dual to a normal vector in both the Jordan and Einstein frames, providing a starting point for establishing relationships for hypersurface geometry between the frames. 

The unit normal vector in the Einstein frame may be defined as
\begin{equation}\label{GCA-DualUnitNormalEF}
   n_\mu := \varepsilon \, \alpha \, \Phi_\mu,
\end{equation}

\noindent where $\varepsilon=\pm 1$ is positive if the unit normal is spacelike and negative if the unit normal is timelike. The normalization factor [equivalent to the lapse function in the $3+1$ formalism] is defined
\begin{equation}\label{GCA-NormalFactor}
   \alpha := \frac{1}{\left(\varepsilon {g}^{\sigma \tau}\Phi_\sigma \Phi_\tau\right)^{1/2} }.
\end{equation}

\noindent In principle, one can choose the foliation so that $\alpha=1$ in the neighborhood of a surface. However, the choice of foliation can be made only once; one may either choose the foliation to set the normalization factor to unity in the Einstein frame ($\alpha=1$) or the Jordan frame ($\hat{\alpha}=1$, where $\hat{\alpha}$ is the Jordan frame counterpart). Setting $\alpha=1$ in the Einstein frame effectively fixes the normalization factor in the Jordan frame and vice versa.

In the Jordan frame, the lowered index normal vector may be written as
\begin{equation}\label{GCA-DualUnitNormalEF2}
   \mathfrak{n}_\mu := \hat{\varepsilon} \, \hat{\alpha} \, \Phi_\mu
   = \aleph \, n_\mu .
\end{equation}

\noindent The quantities $\hat{\varepsilon}$ and $\hat{\alpha}$ are the Jordan frame counterparts to the quantities ${\varepsilon}$ and ${\alpha}$, and 
\begin{equation}\label{GCA-NormRatio}
   \aleph := \frac{\hat{\varepsilon}}{\varepsilon} \frac{\hat{\alpha}}{\alpha} = |\varepsilon \, g^{\mu \nu} \, \mathfrak{n}_\mu \, \mathfrak{n}_\nu |^{1/2} = |\hat{\varepsilon} \, \bar{\mathfrak{g}}^{\mu \nu} \, n_\mu \, n_\nu|^{-1/2} ,
\end{equation}

\noindent where the last two equalities are obtained by contracting \eqref{GCA-DualUnitNormalEF2}.

In the Jordan frame, one must distinguish between the lowered index normal vector $\mathfrak{n}_\nu$ and the raised index normal vector $\bar{\mathfrak{n}}^\mu$, which we write as
\begin{equation}\label{GCA-UnitNormalEF}
   \begin{aligned}
   \bar{\mathfrak{n}}^\mu 
   & := \bar{\mathfrak{g}}^{\mu \nu} \, \mathfrak{n}_\nu \\
   &\,= \Psi^{-1} \, g^{\sigma \tau} \, \bar{A}{^\mu}{_\sigma} \, \bar{A}{^\nu}{_\tau} \, \mathfrak{n}_\nu ,
   \end{aligned}
\end{equation}

\noindent again following the convention that indices are raised and lowered exclusively with the Einstein frame metric $g_{\mu \nu}$.

The induced metric in the Einstein frame may be rewritten in terms of the Jordan frame quantities as
\begin{equation}\label{GCA-InducedMetric}
   \begin{aligned}
   q_{\mu \nu} 
      &= g_{\mu \nu} - \varepsilon \, n_\mu \, n_\nu \\
      &= \Psi^{-1} \, \bar{A}{^\sigma}{_\rho} \, \bar{A}{^\tau}{_\lambda}  \left[ \delta{^\rho}{_\mu} \, \delta{^\lambda}{_\nu} \mathfrak{q}_{\sigma \tau} 
      + 
      \Delta^{\rho \lambda}{}_{\mu \nu} \,
      \mathfrak{n}_{\sigma} \, \mathfrak{n}_{\tau} \right]
      ,
   \end{aligned}
\end{equation}

\noindent where $\mathfrak{q}_{\mu \nu} := \mathfrak{g}_{\mu \nu} - \hat{\varepsilon} \, \mathfrak{n}_\mu \, \mathfrak{n}_\nu$, and
\begin{equation}\label{GCA-CTensor}
   \begin{aligned}
   \Delta^{\rho \lambda}{}_{\sigma \tau}
        &:= \hat{\varepsilon} \, \delta{^\rho}{_\sigma} \, \delta{^\lambda}{_\tau} - \varepsilon \, \aleph^{-2} \, \Psi \, A{_\sigma}{^\rho} \, A{_\tau}{^\lambda} .
   \end{aligned}
\end{equation}

\noindent Similarly, the inverse induced metric may be written as
\begin{equation}\label{GCA-InverseInducedMetric}
   \begin{aligned}
      q^{\mu \nu} 
      &= g^{\mu \nu} - \varepsilon \, n^\mu \, n^\nu \\
      &= \Psi \,  {A}{_\sigma}{^\rho} \, {A}{_\tau}{^\lambda} \left[ \delta{_\rho}{^\mu} \, \delta{_\lambda}{^\nu} \, \bar{\mathfrak{q}}^{\sigma \tau} - \Delta{_{\rho \lambda}}{^{\mu \nu}} \, \bar{\mathfrak{n}}^\sigma \, \bar{\mathfrak{n}}^\tau \right],
   \end{aligned}
\end{equation}

\noindent and the projection operator as
\begin{equation}\label{GCA-ProjOp}
   \begin{aligned}
   q{^\mu}{_\nu} 
        &= \delta{_\nu}{^\mu} - \varepsilon \, n^\mu \, n_\nu \\
        &= \mathfrak{p}{^\mu}{_\nu} + A{_\tau}{^\sigma} \bar{A}{^\mu}{_\lambda} \Delta_\rho{}^{\rho \lambda}{}_{\sigma} \, \bar{\mathfrak{n}}^\tau \, \mathfrak{n}_\nu
      ,
   \end{aligned}
\end{equation}

\noindent where $\mathfrak{p}{^\mu}{_\nu}:=\delta{_\nu}{^\mu} - \hat{\varepsilon} \, \bar{\mathfrak{n}}^\mu \, \mathfrak{n}_\nu$.

The Jordan frame extrinsic curvature $\mathfrak{K}_{\mu \nu} $ is given by the expression
\begin{equation}\label{GCA-ExtrinsicCurvature}
\begin{aligned}
   \mathfrak{K}_{\mu \nu} 
   &= \hat{\nabla}_\mu \mathfrak{n}_\nu - \hat{\varepsilon} \, \mathfrak{n}_{\mu} \, \mathfrak{a}_{\nu} ,
\end{aligned}
\end{equation}

\noindent where the Jordan frame acceleration vector $\mathfrak{a}_{\mu}$ is related to the Einstein frame acceleration vector $a_\mu$ according to (noting that the Jordan frame lapse function satisfies $|\hat{\alpha}|=|\aleph \alpha|$)
\begin{equation}\label{GCA-Accel}
\begin{aligned}
   \mathfrak{a}_{\mu}
   &= -\hat{\varepsilon} \, \mathfrak{p}{^\nu}{_\mu} {\nabla}_\nu \ln(\aleph \alpha) \\
   &= \hat{\varepsilon} \, \varepsilon \, a_\mu + \hat{\varepsilon} \, A{_\tau}{^\sigma} \, \bar{A}{^\nu}{_\lambda} \, \Delta_\rho{}^{\rho \lambda}{}_{\sigma} \, \bar{\mathfrak{n}}^\tau \, \mathfrak{n}_\mu \,  {\nabla}_\nu \alpha - \hat{\varepsilon} \, \mathfrak{p}{^\nu}{_\mu} {\nabla}_\nu \ln \aleph
   .
\end{aligned}
\end{equation}

\noindent The extrinsic curvature tensors in the Einstein and Jordan frames may be related by the following expression
\begin{equation}\label{GCA-ExtrinsicCurvatureII}
\begin{aligned}
\aleph \, K_{\mu \nu}
    &=
    \varepsilon \, \mathfrak{K}_{\mu \nu} - \varepsilon \, W{^\sigma}_{\mu \nu} \mathfrak{n}_\sigma - \varepsilon \left( \mathfrak{n}_\nu \, \delta{_\mu}{^\tau} - \mathfrak{n}_{\mu} \, \mathfrak{p}{^\tau}{_\nu} \right) {\nabla}_\tau \ln \aleph \\
    & \quad
    + \varepsilon \, \hat{\varepsilon} \,  \mathfrak{n}_{\mu} \,\mathfrak{n}_\nu \, A{_\alpha}{^\sigma} \, \bar{A}{^\tau}{_\lambda} \, \Delta_\rho{}^{\rho \lambda}{}_{\sigma} \, \bar{\mathfrak{n}}^\alpha \, {\nabla}_\tau \alpha
    .
\end{aligned}
\end{equation}

\noindent One may choose a gauge such that the Einstein frame metric has the Gaussian normal form, in which the lapse function is unity ($\alpha=1$). The extrinsic curvature then simplifies to
\begin{equation}\label{GCA-ExtrinsicCurvatureIIunitlapse}
\begin{aligned}
\aleph \, K_{\mu \nu}
    &=
    \varepsilon \left[ \mathfrak{K}_{\mu \nu} - W{^\sigma}_{\mu \nu} \mathfrak{n}_\sigma - \left( \mathfrak{n}_\nu \, \delta{_\mu}{^\tau} - \mathfrak{n}_{\mu} \, \mathfrak{p}{^\tau}{_\nu} \right) {\nabla}_\tau \ln \aleph \right].
\end{aligned}
\end{equation}

\noindent One can use Eqs. \eqref{GCA-DualUnitNormalEF2}, \eqref{GCA-InducedMetric}, \eqref{GCA-ProjOp}, and \eqref{GCA-ExtrinsicCurvatureIIunitlapse} to rewrite Einstein frame surface quantities in terms of Jordan frame quantities. 

\subsection{Junction conditions in the Jordan frame} \label{sec:JuncCondGen}
Recalling that the junction condition \eqref{GC:BdyFieldEqsIsrael} implies continuity of the Einstein frame induced metric $\llbracket q^{\mu \nu} \rrbracket = 0$, one obtains the condition [making use of Eq. \eqref{GCA-InverseInducedMetric}]
\begin{equation} \label{GC:BdyFieldEqsIsraelJordanNormal}
   \begin{aligned}
    &  \llbracket \Psi \,  {A}{_\sigma}{^\rho} \, {A}{_\tau}{^\lambda} \left( \delta{_\rho}{^\mu} \, \delta{_\lambda}{^\nu} \, \bar{\mathfrak{q}}^{\sigma \tau} - \Delta{_{\rho \lambda}}{^{\mu \nu}} \, \bar{\mathfrak{n}}^\sigma \, \bar{\mathfrak{n}}^\tau \right) \rrbracket = 0.
   \end{aligned}
\end{equation}

\noindent The difference in the extrinsic curvature across the surface may be written as
\begin{equation}\label{GCA-ExtrinsicCurvatureDiff}
\begin{aligned}
\llbracket K_{\mu \nu} \rrbracket
    &=
    \llbracket \varepsilon  \aleph^{-1} \mathfrak{K}_{\mu \nu} - n_\sigma W{^\sigma}_{\mu \nu} - 2 n_{(\mu} {\nabla}_{\nu)} \ln \aleph \\
    & \quad \quad
    - \hat{\varepsilon} \,  n_{\mu} \, n_{\nu} \bar{\mathfrak{n}}^\tau {\nabla}_\tau \aleph \rrbracket ,
\end{aligned}
\end{equation}

\noindent which may be used in conjunction with $\llbracket q^{\mu \nu} \rrbracket = 0$ and 
Eq. \eqref{GC:BdyFieldEqsMetricTR} to obtain the junction conditions in the Jordan frame. Though Eq. \eqref{GCA-ExtrinsicCurvatureDiff} contains derivatives of the Jordan frame metric and $\aleph$, it remains well defined for discontinuities in these quantities so long as the derivatives of the metric have a well-defined limit on each side of the junction surface. One expects this to be the case as far as the geometry in each side of the junction surface is regular. 

In an appropriate gauge, the junction conditions on the coupling tensor $A{_\mu}{^\alpha}$ and the factor $\Psi$ simplify. Given the junction condition \eqref{GC:BdyFieldEqsIsrael}, one may choose a gauge in which the Einstein frame metric $g_{\mu \nu}$ and its inverse $g^{\mu \nu}$ are continuous across the junction surface. The junction conditions for the Jordan frame metric and inverse are given by
\begin{equation} \label{GC:JordanJuncConds}
   \begin{aligned}
    &  \llbracket \mathfrak{g}_{\mu \nu} \rrbracket = \llbracket \Psi \,  A{_\mu}{^\alpha} \, A{_\nu}{^\beta} \rrbracket g_{\alpha \beta} , \\
    &  \llbracket \bar{\mathfrak{g}}^{\mu \nu} \rrbracket = \llbracket \Psi^{-1} \, \bar{A}{^\mu}{_\alpha} \, \bar{A}{^\nu}{_\beta} \rrbracket g^{\alpha \beta},
   \end{aligned}
\end{equation}

\noindent with the differences in the coupling tensor $A{_\mu}{^\alpha}$ and its inverse $\bar{A}{^\mu}{_\alpha}$ across the junction surface being determined by differences in the energy-momentum tensor $\mathfrak{T}_{\mu \nu}$ by way of the field equations $\mathcal{E}{_\pm^\mu}{_\nu} = 0$ \eqref{GC:BulkFieldEqs}. 

\noindent 

%=======================================================================

%-----------------------------------------------------------------------
%-----------------------------------
%-----------------
%--------
%---
%-
%
%
%-
%---
%--------
%-----------------
%-----------------------------------
%-----------------------------------------------------------------------

%=======================================================================
%-----------------------------------------------------------------------
%
%		MEME MODEL
%
%-----------------------------------------------------------------------
%=======================================================================
\section{THE MEME MODEL}\label{sec:MEMe}

%-----------------------------------------------------------------------
%		SUMMARY
%-----------------------------------------------------------------------
\subsection{Action and field equations}
We now consider a specific instance of a generalized coupling theory, termed the MEMe model, which is defined by the action
\begin{equation}\label{GCA-MEMeAction}
\begin{aligned}
S[\varphi,g^{\cdot\cdot},A{_{\cdot}}{^{\cdot}}]= \int d^4x \biggl\{& \left[R - 2 \, \tilde{\Lambda} \right]\sqrt{-{g}} \\
& 
+ 2 \, \kappa\left(L_{m}[\varphi,\bar{\mathfrak{g}}^{\cdot\cdot}] - \frac{\lambda}{\kappa} \right) \sqrt{|\mathfrak{g}|} \biggr\} ,
\end{aligned}
\end{equation}

\noindent where the Jordan frame metric $\mathfrak{g}_{\alpha \beta}$ is defined as (with $A:=A{_\sigma}{^\sigma}$)
\begin{equation}\label{GCA-JordanMetric}
\mathfrak{g}_{\mu \nu} = e^{(4-A)/2} \,  A{_\mu}{^\alpha} \, A{_\nu}{^\beta} \, g_{\alpha \beta} ,
\end{equation}

\noindent and $\tilde{\Lambda}=\Lambda - \lambda$. Unless stated otherwise, indices are raised and lowered using the metric $g_{\mu \nu}$ and $g^{\mu \nu}$. Defining the parameter
\begin{equation}\label{GCA-qparameter}
q:=\frac{\kappa}{\lambda},
\end{equation}

\noindent the equation of ``motion'' for $A{_\mu}{^\alpha}$ takes the following form
\begin{equation}\label{GCA-ExpFEs}
\begin{aligned}
{A}{_\beta}{^\alpha} - \delta{_\beta}{^\alpha} = q \left[ (1/4) \mathfrak{T} \, {A}{_\beta}{^\alpha} - \mathfrak{T}_{\beta \nu} \, \bar{\mathfrak{g}}^{\alpha \nu} \right],
\end{aligned}
\end{equation}

\noindent where $\mathfrak{T}_{\mu \nu}$ is the energy-momentum tensor defined by the functional derivative of $\int L_{m}[\varphi,\bar{\mathfrak{g}}^{\cdot\cdot}] \sqrt{|\mathfrak{g}|} \, d^4x$, and $\mathfrak{T}:=\bar{\mathfrak{g}}^{\mu \nu}\mathfrak{T}_{\mu \nu}$. The trace of Eq. \eqref{GCA-ExpFEs} implies $A=A{_\sigma}{^\sigma}=4$. The gravitational equations are (setting $A=4$)
\begin{equation}\label{GCA-GEN-GFE}
 G_{\mu \nu} +\left[  \Lambda- \, \lambda\left(1 - |A|\right)\right] \, g_{\mu \nu} = \kappa \, |A| \, \bar{A}{^\alpha}{_\mu} \, \bar{A}{^\beta}{_\nu} \, \mathfrak{T}_{\alpha \beta},
\end{equation}

\noindent where $|A|=\det(A{_\cdot}{^\cdot})$.

As discussed in \cite{Feng:2019dwu}, we note that if $|A|$ vanishes, the gravitational equation becomes
\begin{equation}\label{GCA-GEN-GFE-Crit}
 G_{\mu \nu} \approx \left( \lambda - \Lambda \right) \, g_{\mu \nu}.
\end{equation}

\noindent If $q<0$, then one effectively has a de Sitter vacuum and requiring that $\kappa |q| \ll 1$, $\lambda$ dominates and one has a large cosmological constant. Such behavior is of astrophysical interest since it may permit the existence of gravastarlike black hole mimickers \cite{Mazur:2001fv,Mazur:2004fk} without requiring a false vacuum---as we shall see shortly, a sufficiently high density is all that is required.

%-----------------------------------------------------------------------
%		PERFECT FLUID SOLUTION
%-----------------------------------------------------------------------
\subsection{Perfect fluid solution}
Equation \eqref{GCA-ExpFEs} can be solved exactly for a single perfect fluid. The fluid four-velocities $u^\mu$ are constructed from the gradients of the potentials, so it is appropriate to regard $u_\mu$ to be the metric-independent fluid variables; the energy-momentum tensor for the fluid takes the form
\begin{equation} \label{GCA-EnergyMomentumPerfectFluid}
\mathfrak{T}_{\mu \nu} = \left( \hat \rho + \hat p \right)u_\mu u_\nu + \hat p \> \mathfrak{g}_{\mu \nu},
\end{equation}

\noindent where $\mathfrak{T} := \bar{\mathfrak{g}}^{\mu \nu}\mathfrak{T}_{\mu \nu} =3 \hat p - \hat \rho$. Note that while $\bar{\mathfrak{g}}^{\mu \nu} u_\mu u_\nu = -1$, $g^{\mu \nu} u_\mu u_\nu \neq  -1$ in general. Defining $z := -u_\sigma u^\sigma$, one can rescale the four-velocity
\begin{equation}\label{GCA-UnitFlowField}
U^\mu := {u^\mu}/{\sqrt{z}} ,
\end{equation}

\noindent and it follows that $u_\mu \, u_\nu = z U_\mu \, U_\nu$.

One can show that for a perfect fluid, the following is a solution for \eqref{GCA-ExpFEs}
\begin{equation}\label{GCA-AnsatzRSYform}
   A{_\mu}{^\alpha} = {Y} \, \delta{_\mu}{^\alpha} - 4(1-Y) \, U{_\mu} \, U{^\alpha},
\end{equation}
\begin{equation}\label{GCA-YSol}
   Y = \frac{4 (1 - \hat p \, q)}{4 - q \,  (3 \, \hat p - \hat \rho)}.
\end{equation}

\noindent The determinant of $A{_\mu}{^\alpha}$ may be written as
\begin{equation}\label{GCA-ExpAdetYform}
   |A| = Y^3(4-3Y) = \frac{256 \, (1 - \hat p \, q)^3 (q \, \hat \rho + 1)}{[4 - q \, (3 \hat p - \hat \rho) ]^4}.
\end{equation}

\noindent Note that for $q<0$, $|A|$ vanishes at a critical density $|q| \hat{\rho} = 1$, and that for $q>0$, $|A|$ vanishes at a critical pressure $q \hat{p} = 1$. The Jordan metric takes the form
\begin{equation}\label{GCA-TransformedJordanMetric}
\mathfrak{g}_{\mu \nu} = Y^2 g_{\mu \nu} - 8 (Y-2) (Y-1) U_{\mu} U_{\nu}.
\end{equation}

\noindent From Eqs. \eqref{GCA-YSol} and \eqref{GCA-TransformedJordanMetric}, one can see explicitly that discontinuities in the fluid density $\hat \rho$ and/or pressure $\hat p$ will generally produce discontinuities in $Y$ and the metric (though we will discuss one exception below).

The gravitational equation \eqref{GCA-GEN-GFE} can be written in the form
\begin{equation}
G_{\mu \nu}=\kappa \, T_{\mu \nu},
\end{equation}
where $T_{\mu \nu}$ is the effective energy-momentum tensor defined by
\begin{equation}\label{GCA-ExpTmnEffDecomp}
T_{\mu \nu} = \left( \rho + p \right) U_\mu \, U_\nu + p \, g_{\mu \nu} ,
\end{equation}

\noindent and the effective density $\rho$ and pressure $p$ have the form:
\begin{equation}\label{GCA-ExpTmnEffDecompDensityPressure}
\begin{aligned}
\rho  & = |A| \, (\hat p + \hat \rho) - p, \\
p     & = \frac{|A| \, (\hat p \, q - 1) + 1}{q}-\frac{\Lambda}{\kappa}.
\end{aligned}
\end{equation}

\noindent In terms of $\rho$ and $p$, one may write
\begin{equation}\label{GCA-YSoleff}
   Y = \frac{4 (\vartheta_0 - p \, q)}{4 \, \vartheta_0 - q \,  (3 \, p - \rho )} ,
\end{equation}
where $\vartheta_0$ is the following constant
\begin{equation}\label{GCA-vartheta}
\vartheta_0 = 1  - \Lambda/\lambda .
\end{equation}

More generally, one can show that for a transformation of the form (with $\Phi$ being an arbitrary factor), 
\begin{equation}\label{GCA-EnPress}
\begin{aligned}
\rho^\prime &  =  \left[ \frac{\Phi (1 + q \, \rho) - 1}{q}\right] + \Xi,
\\
p^\prime & = \left[ \frac{\Phi \, (p \, q - 1) + 1}{q} \right]- \Xi,
\\
\vartheta^\prime & = \Phi \left( \vartheta - 1 \right) + 1 - q \, \Xi ,
\end{aligned}
\end{equation}

\noindent the functional form of the following quantity is preserved
\begin{equation}\label{GCA-YSol2}
Y = \frac{4 (\vartheta - q \, p)}{4 \vartheta - q \left(3 p - \rho\right)} = \frac{4 (\vartheta^\prime - q \, p^\prime)}{4 \vartheta^\prime - q \left(3 p^\prime - \rho^\prime\right)} = Y^\prime.
\end{equation}

\noindent One recovers the transformation \eqref{GCA-ExpTmnEffDecompDensityPressure} by
\begin{equation}\label{GCA-Transformation}
   \begin{aligned}
   \Phi = |A| , 
   \qquad 
   \Xi = \frac{\Lambda}{q \, \lambda} ,
   \qquad
   \vartheta = 1.
   \end{aligned}
\end{equation}

%-----------------------------------------------------------------------
%		JORDAN FRAME DUST
%-----------------------------------------------------------------------
\subsection{Jordan frame dust}
For later use, we obtain the equation of state for an effective fluid in the Einstein frame for a Jordan frame dust fluid. For simplicity, consider the case where $\Lambda=0$. For the Jordan frame dust fluid, one has $\hat \rho = 0$ so that under the transformations \eqref{GCA-ExpTmnEffDecompDensityPressure} and \eqref{GCA-vartheta}, one has in the Einstein frame
\begin{equation}\label{GCA-EnPressDust}
   \begin{aligned}
   \rho = \frac{|A|\, (1 + q \, \hat \rho) - 1}{q} ,
   \quad
   p  = \frac{1 - |A|}{q} ,
   \quad
   \vartheta_0  = 1 .
   \end{aligned}
\end{equation}

\noindent The second expression yields $|A| = 1 - q \, p$. One uses these expressions and \eqref{GCA-YSoleff} to relate $|A|$ and $Y$:
\begin{equation}\label{GCA-Y}
   Y = \frac{4 \, |A|}{1 + 3 \, |A| + q \, \rho} , 
    \qquad
   q \rho
   = \frac{|A|^2}{Y^4} - 1 ,
\end{equation}

\noindent where we have used $|A|=Y^3(4-3Y)$ [see \eqref{GCA-ExpAdetYform}] to derive the second relation from the first one. The above may be rewritten
\begin{equation}\label{GCA-PhiRho}
   16 \sqrt{ 1 + q \, \rho }
   = \frac{\left(1 + 3 \, |A| + q \, \rho\right)^2}{|A|},
\end{equation}

\noindent which may be solved for $|A|$ to obtain
\begin{equation}\label{GCA-PhiRho2}
   |A|
   =
   \frac{X}{9} \left[ 8 + 4 \sqrt{4 - 3 X} - 3 X \right],
\end{equation}

\noindent where $X=\sqrt{1 + q \, \rho}$. Combined with the expression $|A| = 1 - q \, p$, Eq. \eqref{GCA-PhiRho2} yields an equation of state $p(\rho)$
\begin{equation}\label{GCA-EoS}
   p = \frac{1}{q} + \frac{X}{9 \, q} \left[3 X - 8 - 4 \sqrt{4 - 3 X} \right] .
\end{equation}

\noindent For $q \, \rho \ll 1$, the effective sound speed is given by
\begin{equation}\label{GCA-ESS}
   v_e^2 := \frac{\partial p}{\partial \rho} = \frac{3}{4} \, q \, \rho + \mathcal{O}( [q \, \rho]^2 ) .
\end{equation}

%=======================================================================

%-----------------------------------------------------------------------
%		SINGULAR HYPERSURFACES IN THE MEMe MODEL
%-----------------------------------------------------------------------
\subsection{Singular hypersurfaces and junctions in the MEMe model}
In the MEMe model, the junction conditions of Sec. \ref{sec:JuncCondGen} may be written in a simplified form for a single fluid under a slight (but physically reasonable) restriction. Here, we assume also the Gaussian normal gauge in the Einstein frame, as well as the continuity of the Einstein frame metric $g_{\mu \nu}$ and fluid four-velocity $U^\mu$. The lapse ratio $\aleph$ may be written as
\begin{equation}\label{GCA-Xisq}
\aleph^2 = \frac{\hat{\varepsilon } (4-3 Y)^2 Y^2}{\varepsilon  (4-3 Y)^2 + 8 \, U_{\perp}^2 (Y-2) (Y-1)},
\end{equation}

\noindent and the raised Jordan frame normal vector takes the form
\begin{equation}\label{GCA-nju}
\bar{\mathfrak{n}}^\mu =
    \frac{\aleph  ( (4-3 Y)^2 n^{\mu} + 8 U_{\perp} (Y-2) (Y-1) U^{\mu})}{\epsilon  (4 - 3 Y)^2 Y^2},
\end{equation}

\noindent where 
\begin{equation}\label{GCA-OrthCond}
U_{\perp}:=n_\mu \, U^\mu.
\end{equation}

\noindent The quantity $U_{\perp}$ vanishes when the fluid four-velocity is tangent to the junction hypersurface. This latter condition is equivalent to its counterpart in the Jordan frame; it is straightforward to show that $U_{\perp}=0$ if $\bar{\mathfrak{g}}^{\mu \nu} \mathfrak{n}_\mu u_\nu = 0$. Physically, this corresponds to the condition that no fluid passes into or out of the hypersurface.

In the limit $U_{\perp}=0$, and assuming $\varepsilon=\hat{\varepsilon}=1$ (which specifies no signature change of the normal vector between spacelike Einstein and Jordan frame normal vectors), one has $\aleph^2=Y^2$. In this same limit, we note that the raised index Jordan frame normal vector simplifies to $\bar{\mathfrak{n}}^\mu = Y^{-1} \, n^{\mu}$. When $U_{\perp}=0$, the Jordan frame projection tensor coincides with the Einstein frame projection tensor
\begin{equation}\label{GCA-ProjectionTensorMEMe}
    \mathfrak{p}_{\mu}{}^{\nu} = q_{\mu}{}^{\nu} .
\end{equation}

\noindent The induced metric for $U_{\perp}=0$ has the form 
\begin{equation}\label{GCA-InducedMEMe}
    \mathfrak{q}_{\mu \nu} = (2 Y - 1)  \, q_{\mu \nu}  + 8 (Y-1) U_{\mu} U_{\nu},
\end{equation}

\noindent and the inverse induced metric becomes
\begin{equation}\label{GCA-InvInducedMEMe}
    \mathfrak{q}^{\mu \nu} = \frac{q^{\mu \nu}}{Y^2} + \frac{8 \epsilon  U^{\mu} U^{\nu} (Y-2) (Y-1)}{Y^2 (4-3 Y)^2}.
\end{equation}

\noindent The junction condition \eqref{GC:BdyFieldEqsIsrael} implies continuity of ${q}_{\mu \nu}$ and ${q}^{\mu \nu}$, which leads to the Jordan frame expressions:
\begin{equation}\label{GCA-InvInducedMEMeJunction}
    \begin{aligned}
    \left\llbracket Y^2 \mathfrak{q}^{\mu \nu} \right\rrbracket 
    &= 
    \left\llbracket \frac{8 \epsilon (Y-2) (Y-1)}{(4-3 Y)^2} \right\rrbracket  U^{\mu} U^{\nu}, \\
    \left\llbracket \frac{\mathfrak{q}_{\mu \nu}}{(2 Y - 1)} \right\rrbracket 
    &=   
    \left\llbracket \frac{8 (Y-1)}{(2 Y - 1)} \right\rrbracket  U_{\mu} U_{\nu}.
    \end{aligned}
\end{equation}

\noindent Note that we have adopted the Gaussian normal gauge in the Einstein frame and assumed that $g_{\mu\nu}$ and $U^{\mu}$ are continuous across the junction hypersurface. We therefore find that the discontinuity in the induced metric is proportional to the outer product of the fluid-four velocity. 

The extrinsic curvature relation \eqref{GCA-ExtrinsicCurvatureIIunitlapse} in this limit becomes,
\begin{equation}\label{GCA-ExtrinsicCurvMEMe}
    \begin{aligned}
        Y^2 K_{\mu \nu} = \, & Y \mathfrak{K}_{\mu \nu} 
            - 16 (Y-2) (Y-1) n^{\sigma} U_{(\mu} \nabla _{[\nu)}U_{\sigma]} \\
            & - (Y q_{\mu \nu} + 4 (3 - 2 Y) U_{\mu} U_{\nu} ) n^{\sigma} \nabla _{\sigma}Y .
    \end{aligned}
\end{equation}

\noindent One may then rewrite the junction condition \eqref{GC:BdyFieldEqsMetricTR} as the following,
\begin{equation}\label{GCA-ExtrinsicCurvDiffMEMe}
\begin{aligned}
    \left\llbracket \frac{\mathfrak{K}_{\mu \nu}}{Y} \right\rrbracket
    = & \, \hat{S}_1 \, n^{\alpha} U_{(\mu} \nabla _{[\nu)}U_{\alpha]} + \hat{S}_2 \,  U_{\mu} U_{\nu} + \hat{S}_3 \, q_{\mu \nu} \\
    & + \kappa \left\llbracket S_{\mu \nu} - \frac{1}{2} S \, q_{\mu \nu} \right\rrbracket,
\end{aligned}
\end{equation}

\noindent where the following scalar quantities are defined as
\begin{equation}\label{GCA-ExtrinsicCurvDiffMEMeScalarCoeffs}
\begin{aligned}
    \hat{S}_1 :=\, &
    \left\llbracket \frac{16 (Y-2) (Y-1) }{Y^2}  \right\rrbracket , \\
    \hat{S}_2 :=\, &
    \left\llbracket \frac{4 (3 - 2 Y)  n^{\sigma} \nabla _{\sigma} Y}{Y^2} \right\rrbracket , \\
    \hat{S}_3 :=\, &
    \left\llbracket \frac{n^{\sigma} \nabla _{\sigma}Y}{Y^2} \right\rrbracket .
\end{aligned}
\end{equation}

\noindent Equations \eqref{GCA-ExtrinsicCurvDiffMEMe} and \eqref{GCA-ExtrinsicCurvDiffMEMeScalarCoeffs} establish the junction conditions for the MEMe model in the Jordan frame, under the assumption that the fluid does not cross the junction hypersurface and that the hypersurface is timelike in both the Einstein and Jordan frames. We note that the term containing the factor $\hat{S}_1$ vanishes in the absence of vorticity in $U^\mu$. The respective quantities $\hat{S}_2-\hat{S}_3$ and $\hat{S}_3$ may then be thought of as surface energy and stresses, which are both dependent on the normal derivative of $Y$. As we remarked in the general case, the derivatives of discontinuous quantities (in this case, the normal derivative of $Y$) may be of concern. However, if $n^{\sigma} \nabla _{\sigma}Y$ has a well-defined limit on each side of the junction hypersurface, the quantities $\hat{S}_2$ and $\hat{S}_3$ remain well defined, and (assuming that no additional constraints are introduced at the surface) there is nothing particularly pathological about Eqs. \eqref{GCA-ExtrinsicCurvDiffMEMe} and \eqref{GCA-ExtrinsicCurvDiffMEMeScalarCoeffs}; they provide a rather straightforward correspondence between the Einstein and Jordan frame junction conditions in the MEMe model (keeping in mind the aforementioned assumptions).

In general, discontinuities in the fluid density and the pressure will induce discontinuities in $Y$, so that the quantities $\hat{S}_1$, $\hat{S}_1$, and $\hat{S}_1$ contribute nontrivially to the junction condition \eqref{GCA-ExtrinsicCurvDiffMEMe}. However, we wish to point out here one exception. Recall the transformation \eqref{GCA-ExpTmnEffDecompDensityPressure}, setting $\Xi=1-\vartheta=0$ [corresponding to $\Lambda=0$ and thus $\vartheta_0=1$ in \eqref{GCA-ExpTmnEffDecompDensityPressure} and \eqref{GCA-vartheta}] for simplicity. Now consider a surface located at some coordinate value $r=r_0$, and assume a uniform density $\hat \rho_0$ and pressure $\hat p_0$ for $r<r_0$. A general $Y$-preserving transformation of the density and pressure of the form in \eqref{GCA-EnPress} (applied to the Jordan frame quantities) may be written as
\begin{equation}
   \begin{aligned}
   \hat{\rho}' &  
   = \hat \rho_0 + (1 + q \, \hat \rho_0 ) \, c ,
   \\
   \hat{p}' & 
   = \hat p_0 - (1 - q \, \hat p_0 ) \, c ,
   \end{aligned}
\end{equation}

\noindent where $c=(\Phi-1)/q$ is an arbitrary factor. Considering $\hat{\rho}'$ and $\hat{p}'$ as the Jordan frame energy density and pressure in the region $r>r_0$ or inside the thin layer corresponding to the junction hypersurface, this transformation provides a condition for both the Einstein and Jordan frame metrics to be continuous. As for the jumps, $\hat{\rho}$ and $\hat{p}$ satisfying the following expression are permitted by this condition
\begin{equation}
\frac{\llbracket \hat p \rrbracket}{\llbracket \hat \rho \rrbracket}
= 
-\frac{1 - q \, \hat p_0}{1 + q \, \hat \rho_0} .
\end{equation}

\noindent As for singular hypersurfaces in the absence of other matter (assuming $U^\mu \nabla_\mu r = u^\mu \nabla_\mu r = 0$), both metrics are continuous when the comoving surface energy density $\hat \rho_s$ and isotropic surface stress $\hat \sigma_s$ satisfy $\hat \rho_s = -\hat \sigma_s$.

%=======================================================================

%-----------------------------------------------------------------------
%-----------------------------------
%-----------------
%--------
%---
%-
%
%
%-
%---
%--------
%-----------------
%-----------------------------------
%-----------------------------------------------------------------------

%=======================================================================
%-----------------------------------------------------------------------
%
%		SHARP GRADIENTS
%
%-----------------------------------------------------------------------
%=======================================================================
\section{Sharp gradients}\label{sec:SGrad}

%=======================================================================
%		SHARP GRADIENTS: STATIC CASE
%=======================================================================
\subsection{Static case}
In general, we have seen that in the MEMe model, the Jordan frame metric tensor is not continuous in the presence of sharp boundaries and/or singular hypersurfaces unless the pressure and density satisfy certain constraints at the surface. However, the junction conditions seem to indicate that no pathologies appear so long as the derivatives of the discontinuous quantities have well-defined limits on each side of the hypersurface. 

One might wonder whether the formalism developed thus far suffices. In particular, discontinuities in the metric components may be regarded as approximating sharp gradients. One might wonder if the backreaction from sharp gradients in the metric contribute significantly to the behavior of the fluid.

\begin{figure}[!htp]
  \includegraphics[
  width=\columnwidth
  ]{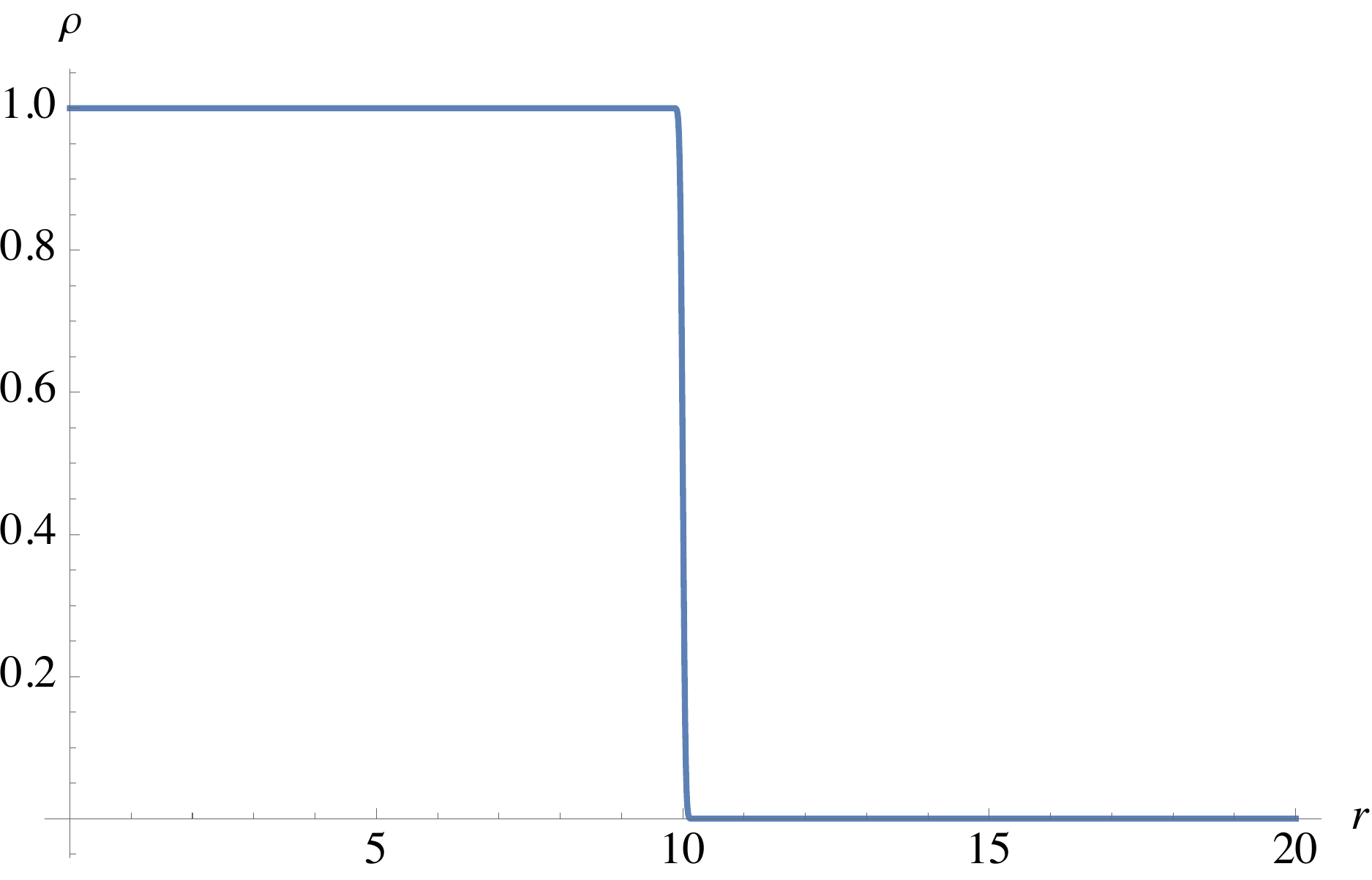}
  \caption{
    Density profile given by Eq. \eqref{rhoErf} for the parameters $R_0=10$, $\sigma=1/10$, and $m_0=1$, with $\rho_0 = 4 \pi \, m_0 /R_0^3$. The region near $R_0=10$ is plotted here, since away from $R_0$, $\rho$ is approximately constant.
  }
  \label{fig:denspfu}
\end{figure}

To examine this question, we consider a fluid with a given density profile containing a sharp gradient and solve the Tolman-Oppenheimer-Volkoff (TOV) equation in the Einstein frame. By way of the transformations presented earlier, one may obtain the corresponding profiles in the Jordan frame. We consider here a static, spherically symmetric geometry, described by a line element of the form,
    \begin{equation} \label{SpLE}
        ds^2 =  - f(r) \, dt^2 
                + h(r) \, dr^2 +  r^2 \, d\Omega^2 ,
    \end{equation}

\noindent where $d\Omega^2 = d\theta^2 + \sin^2(\theta)\, d\phi^2$ is the line element for the 2-sphere. Since the metric is diagonal, and the system is static, then from Eq. \eqref{GCA-TransformedJordanMetric} it follows that if the four-velocity $U^\mu$ is proportional to $\partial/\partial t$ (as one requires from staticity), then the spatial part of the metric is rescaled by a factor of $Y^2$. The areal radius $r$ in the Einstein frame is then related to the areal radius $\mathfrak{r}$ in the Jordan frame by
    \begin{equation} \label{ArealRadii}
        \mathfrak{r} = Y \, r .
    \end{equation}
\noindent We consider a spherically symmetric density profile of the following form in the Einstein frame (with $r$ being the areal radius)
\begin{equation} \label{rhoErf}
\rho = \frac{1}{2} \rho_0 \left[1-\text{erf}\left(\frac{(r-R_0) (R_0+r)}{r \sigma }\right)\right] ,
\end{equation}

\noindent which is mostly constant for $r<R_0$ and mostly zero for $r>R_0$, with $r$ being areal radius, and $\rho_0 = 4 \pi \, m_0 /R_0^3$. The density profile near $R_0$ is plotted in Fig. \ref{fig:denspfu}. The derivative of $\rho$ is controlled by the boundary thickness parameter $\sigma$,
\begin{equation} \label{GradrhoErf}
\left.\frac{d\rho}{dr}\right|_{r=R_0} = -\frac{2 \rho_0}{\sqrt{\pi } \sigma }.
\end{equation}

\noindent To obtain the corresponding pressure profile in the Einstein frame, one may solve the (TOV) equation
\begin{equation} \label{TOV}
\frac{dp}{dr} = -(\rho + p) \left[\frac{m(r)+4 \pi r^3 p}{(r-2m(r))r}\right] ,
\end{equation}

\noindent where $m(r)$ is given by
\begin{equation} \label{massfunc}
m(r) = \int_0^r \rho(\tilde r) \tilde r^2 d\tilde r.
\end{equation}

\noindent Since the density in \eqref{rhoErf} is approximately uniform, the central pressure is given by
\begin{equation} \label{CentralPressure}
p_c \approx \rho_0 \left[\frac{1-\sqrt{1-{2 m_0}/{R_0}}}{3 \sqrt{1-{2 m_0}/{R_0}}-1}\right] .
\end{equation}

\begin{figure}[]
\subfloat[\label{fig:massfua}]{%
\includegraphics[width=\columnwidth]{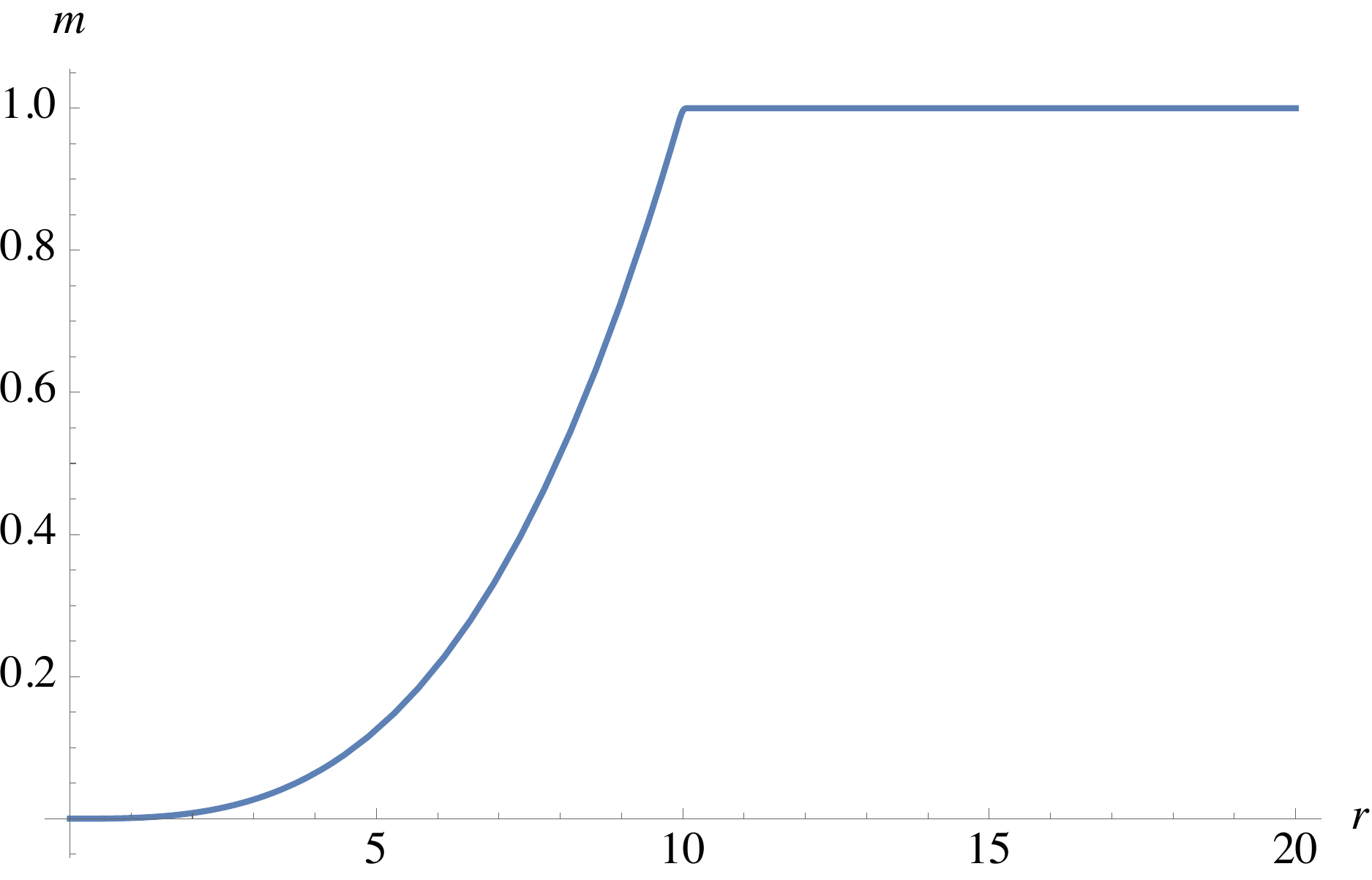}%
}\\
\subfloat[\label{fig:massfub}]{%
\includegraphics[width=\columnwidth]{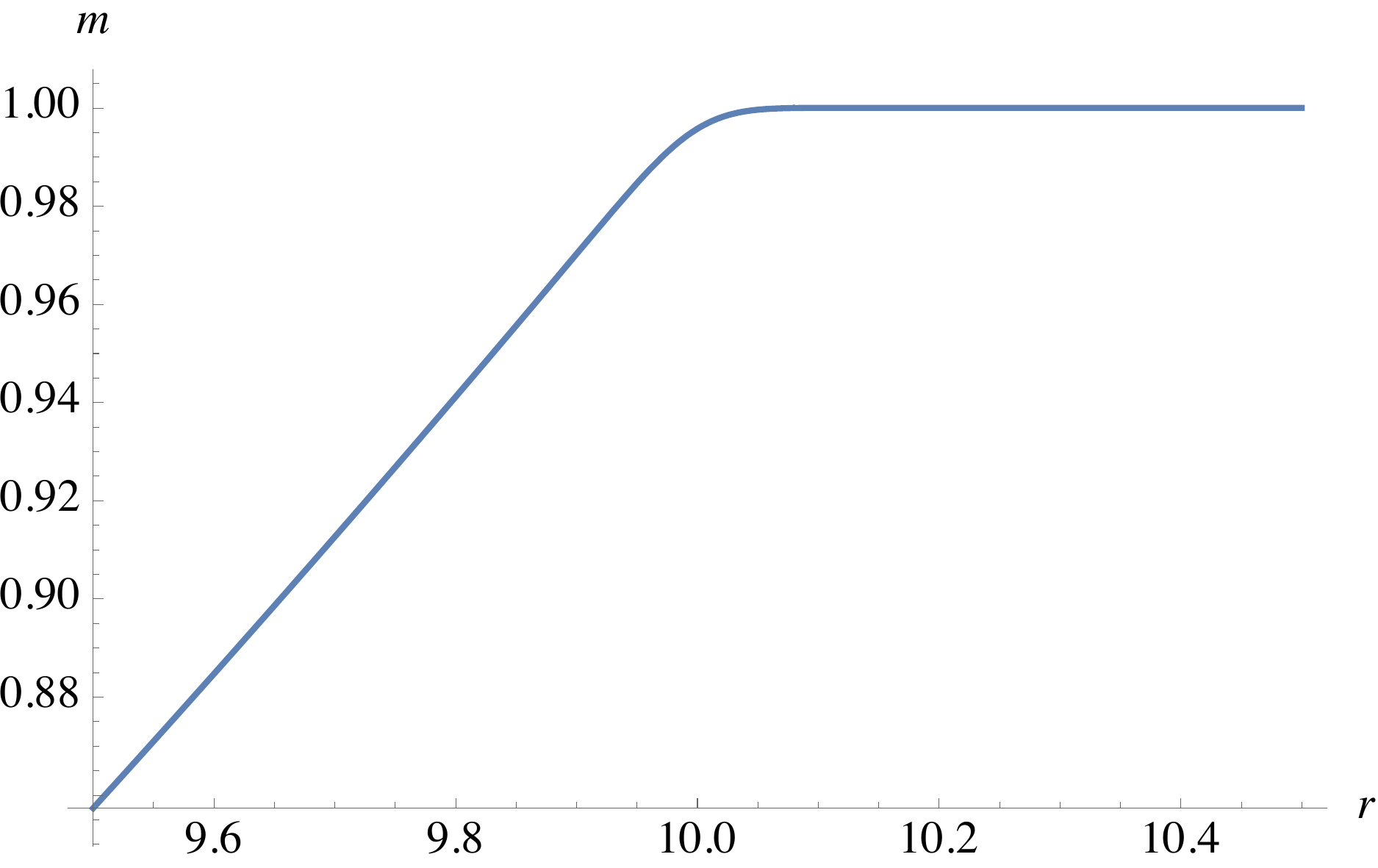}%
}
\caption{
Mass function for the parameters $R_0=10$, $\sigma=1/10$, and $m_0=1$. (a) is the profile in the domain $[0,20]$ and (b) is the profile in the domain $[9.5,10.5]$.
}
\label{fig:massfu}
\end{figure}

\begin{figure}[!ht]
\subfloat[\label{fig:presssua}]{%
\includegraphics[width=\columnwidth]{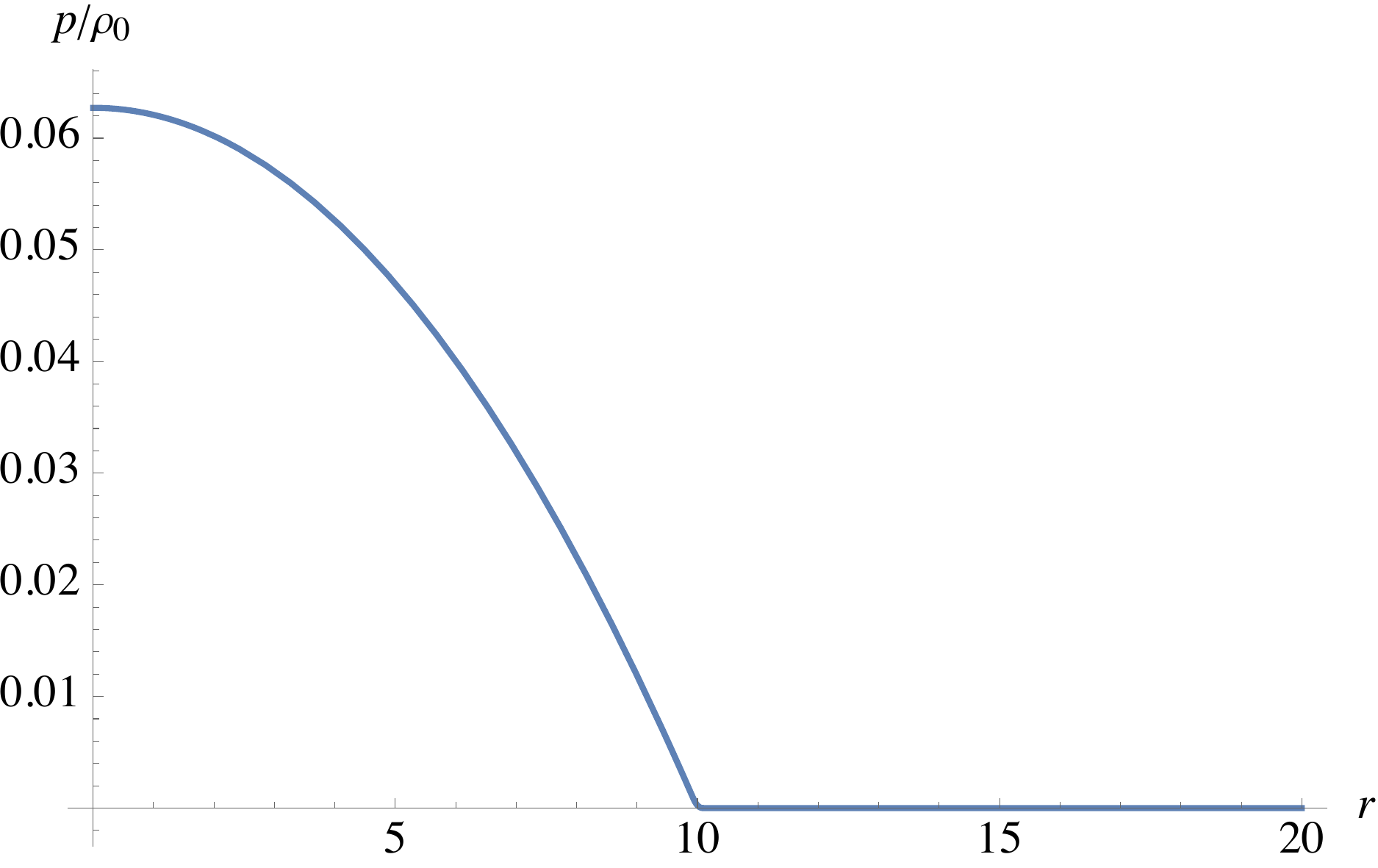}%
}\\
\subfloat[\label{fig:presssub}]{%
\includegraphics[width=\columnwidth]{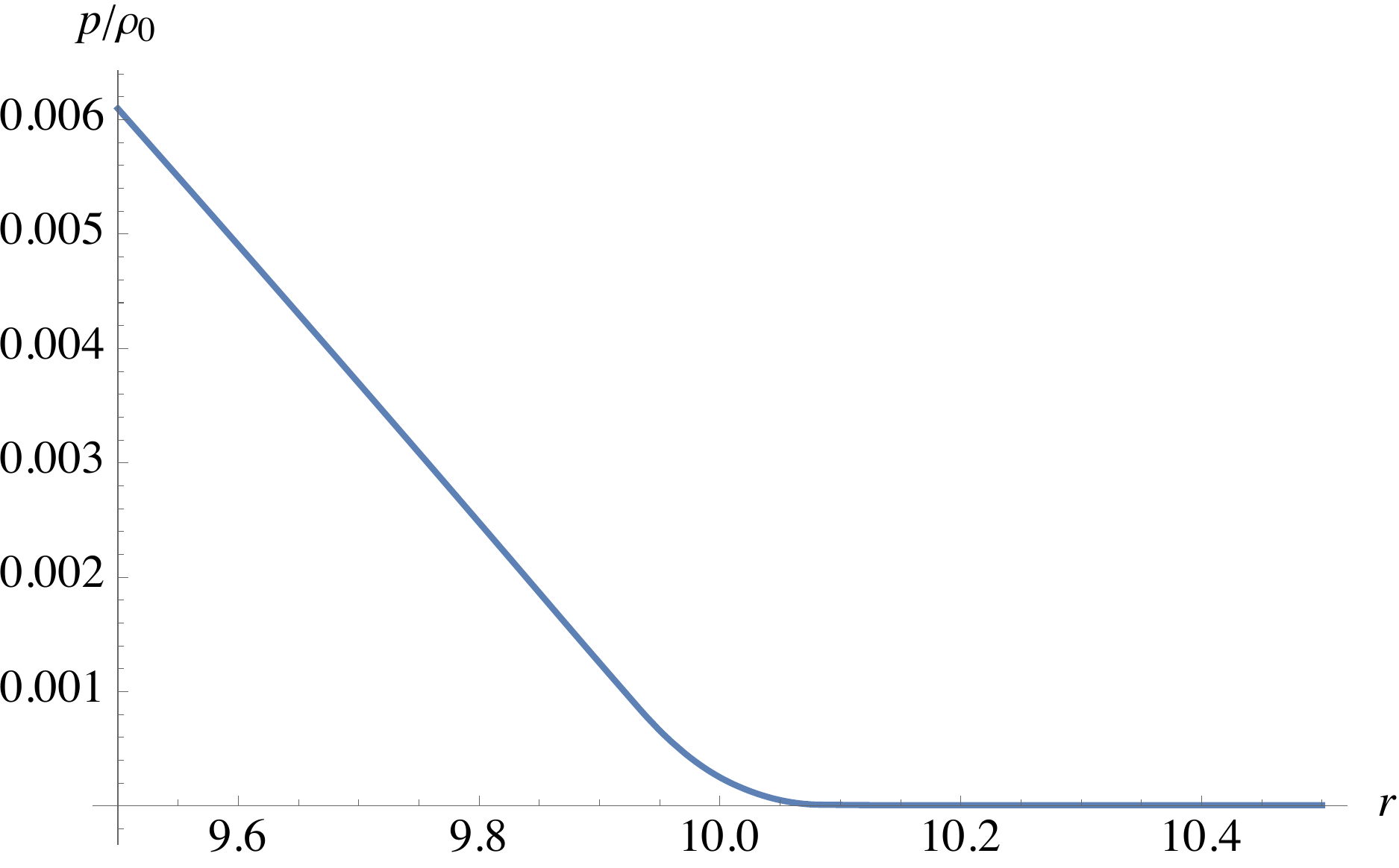}%
}
\caption{
Pressure for the parameters $R_0=10$, $\sigma=1/10$, and $m_0=1$. (a) is the profile in the domain $[0,20]$ and (b) is the profile in the domain $[9.5,10.5]$.
}
\label{fig:presssu}
\end{figure}

\noindent We solve Eqs. \eqref{TOV} and \eqref{massfunc} for the pressure and mass profiles numerically, with the parameters $R_0=10$, $\sigma=1/10$, and $m_0=1$, with the density $\rho_0 = 4 \pi \, m_0 /R_0^3$. The results are presented in Figs. \ref{fig:massfu} and \ref{fig:presssu}.

\begin{figure}[!htp]
  \subfloat[\label{fig:diffdensitya}]{%
    \includegraphics[width=\columnwidth]{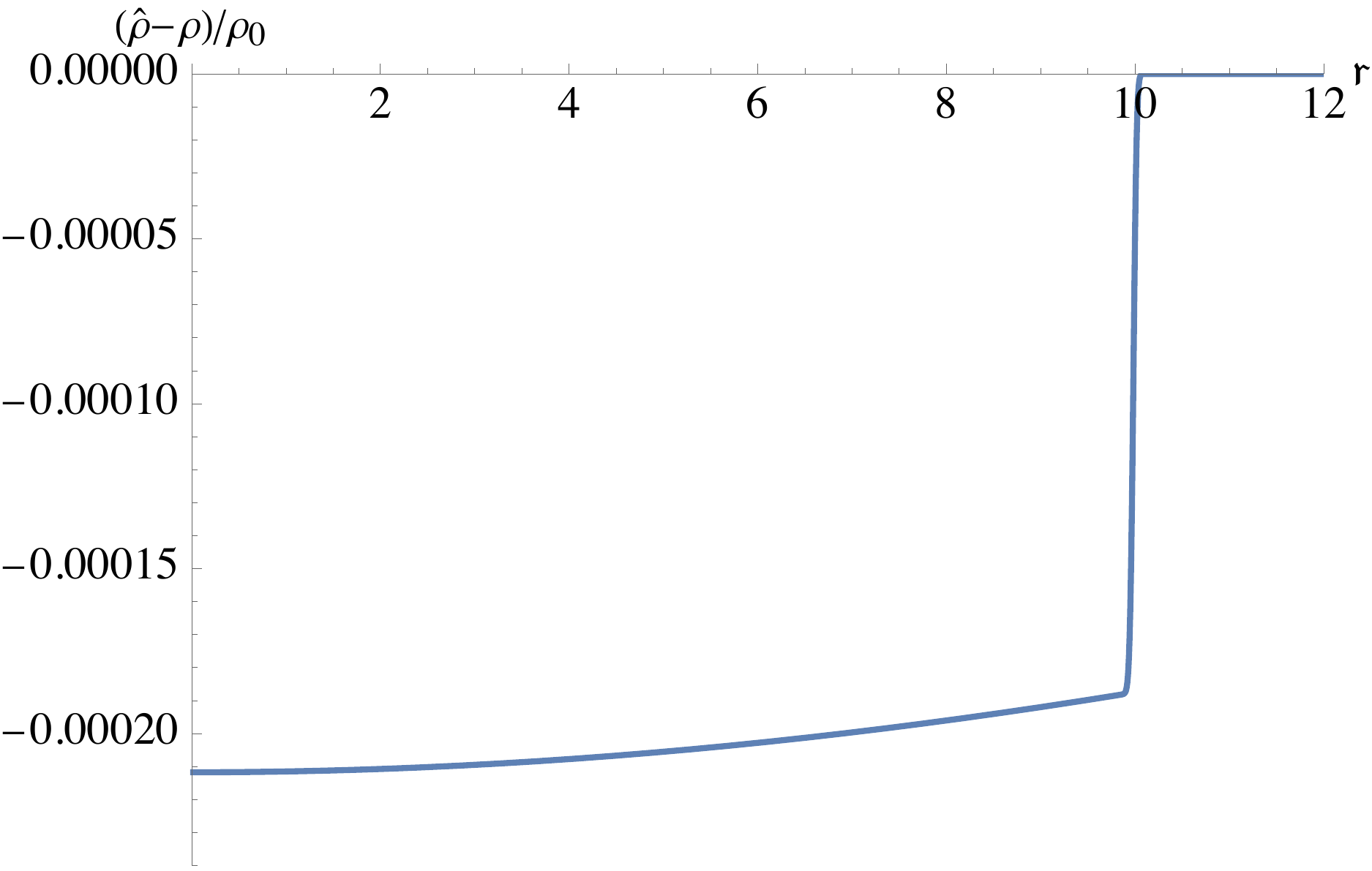}%
    }\\
  \subfloat[\label{fig:diffdensityb}]{%
    \includegraphics[width=\columnwidth]{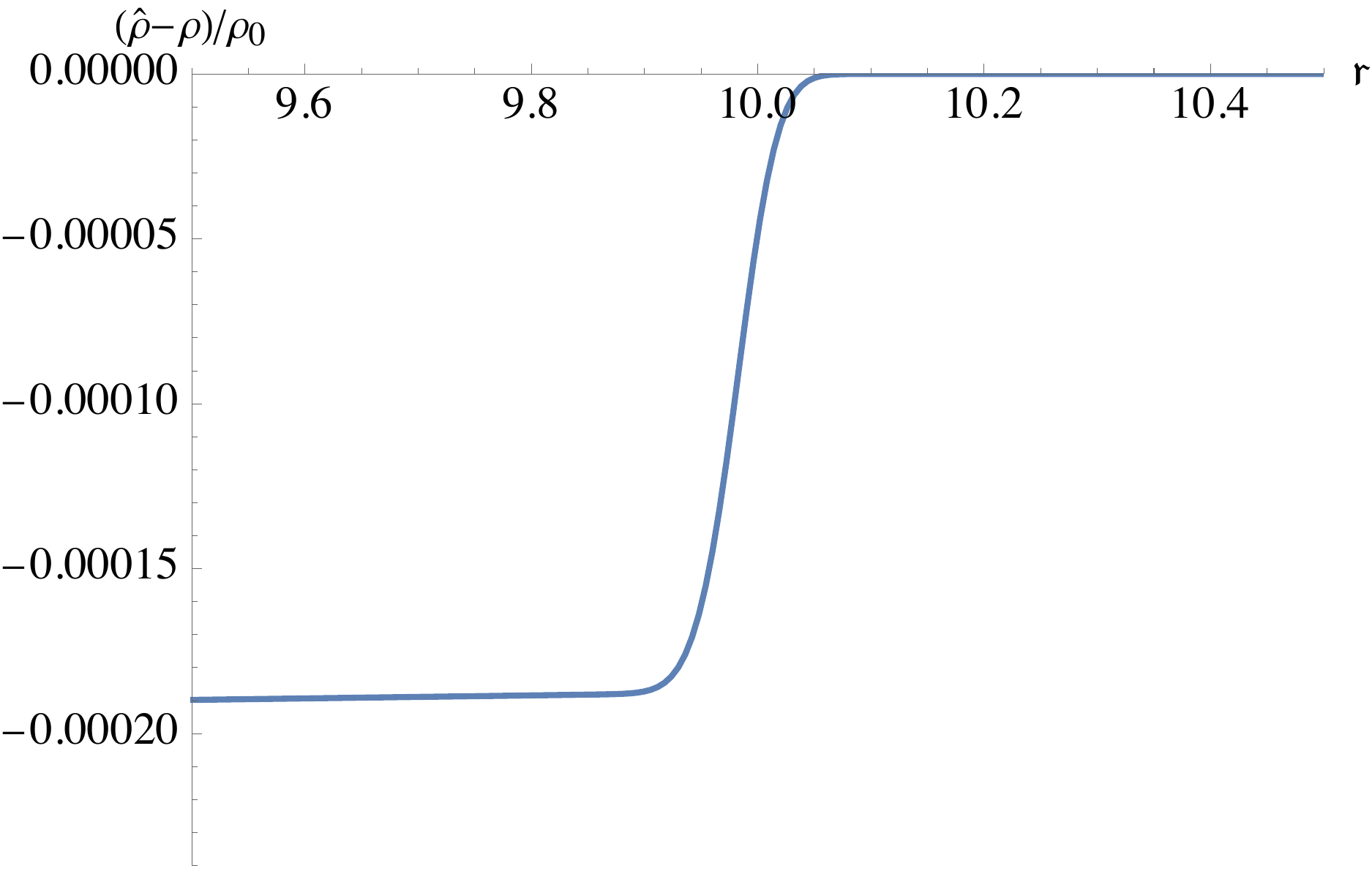}%
    }
  \caption{
    Differences between Jordan and Einstein frame density profiles for the parameters $R_0=10$, $\sigma=1/10$, $m_0=1$, and $q=-2\pi/3$, plotted with respect to the Jordan frame areal radius $\mathfrak{r}$. (a) is the profile in the domain $[0,20]$ and (b) is the profile in the domain $[9.5,10.5]$.
  }
  \label{fig:diffdensity}
\end{figure}

\begin{figure}[!htp]
  \subfloat[\label{fig:diffpressa}]{%
    \includegraphics[width=\columnwidth]{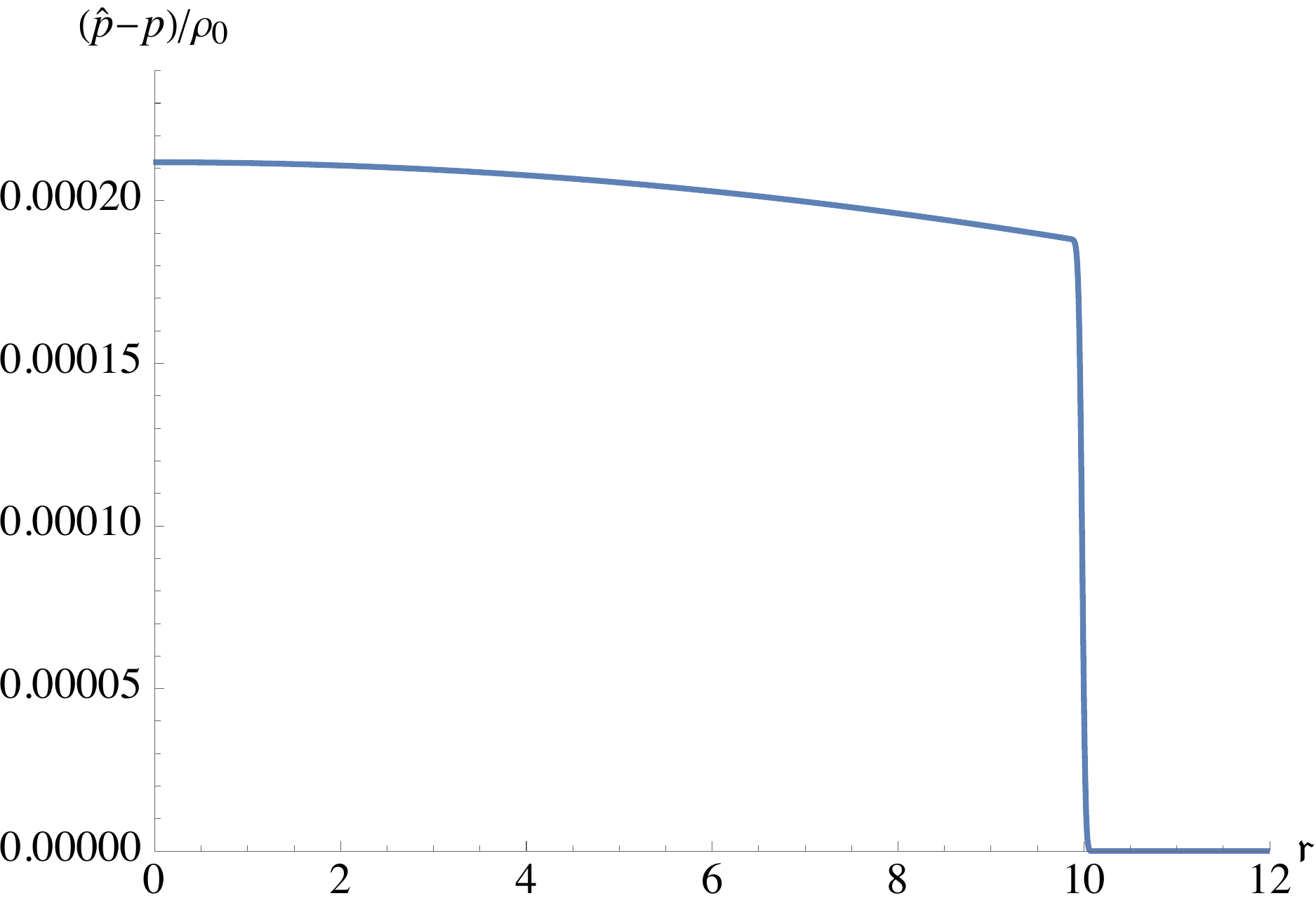}%
    }\\
  \subfloat[\label{fig:diffpressb}]{%
    \includegraphics[width=\columnwidth]{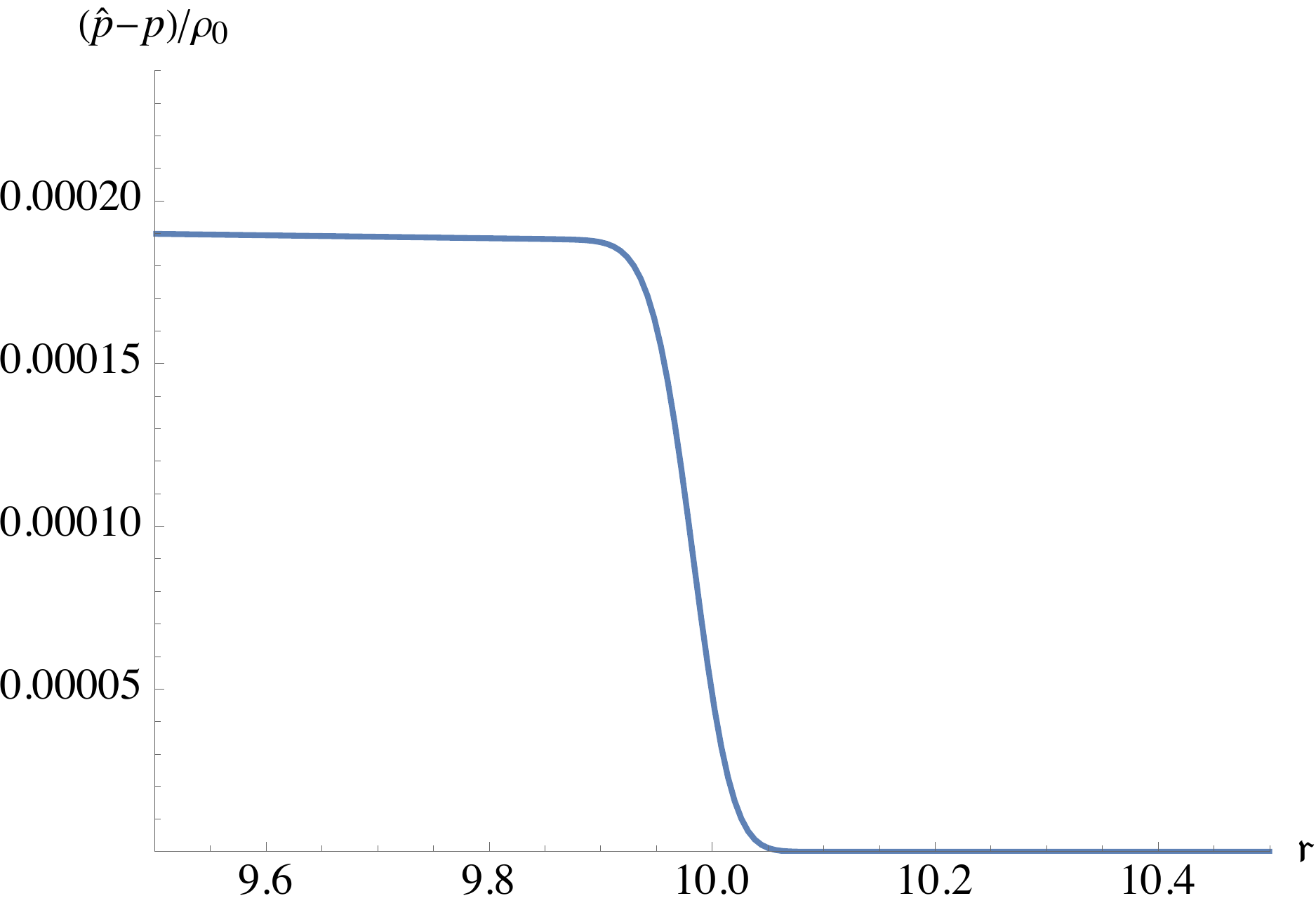}%
    }
  \caption{
    Differences between Jordan and Einstein frame pressure profiles for the parameters $R_0=10$, $\sigma=1/10$, $m_0=1$, and $q=-2\pi/3$, plotted with respect to the Jordan frame areal radius $\mathfrak{r}$. (a) is the profile in the domain $[0,20]$ and (b) is the profile in the domain $[9.5,10.5]$.
  }
  \label{fig:diffpressure}
\end{figure}

Of course, $\rho$ and $p$ are Einstein frame quantities. The corresponding Jordan frame quantities may be obtained from Eqs. \eqref{GCA-YSol}, \eqref{GCA-ExpAdetYform}, \eqref{GCA-ExpTmnEffDecompDensityPressure}, and the areal radius $\mathfrak{r}$ in the Jordan frame is given by Eq. \eqref{ArealRadii}. For small $q$, $\mathfrak{r} \approx r$. We choose $q=-2\pi/3 \approx -2.094$, which corresponds to setting $q \hat \rho_0 = -5 \times 10^{-4}$, and plot the differences between Jordan frame and Einstein frame profiles in Figs. \ref{fig:diffdensity} and \ref{fig:diffpressure} with respect to the Jordan frame areal radius $\mathfrak{r}$. We find that the difference profiles for the density and pressure in Figs. \ref{fig:diffdensity} and \ref{fig:diffpressure}  are flipped with respect to each other, and have steep (but smooth) gradients near $\mathfrak{r} \approx 10$. Here, we see that the differences between the Jordan frame densities and pressures are small compared to their Einstein frame counterparts (note also that the Jordan frame pressure is higher than the Einstein frame pressure), so the Jordan frame does not introduce a violation of energy conditions. We also find that the differences in Einstein and Jordan frame density and pressure profiles are roughly on the order of $q \hat \rho$, as one might expect.

\begin{figure}[!htp]
  \subfloat[\label{fig:Ystatica}]{%
    \includegraphics[width=\columnwidth]{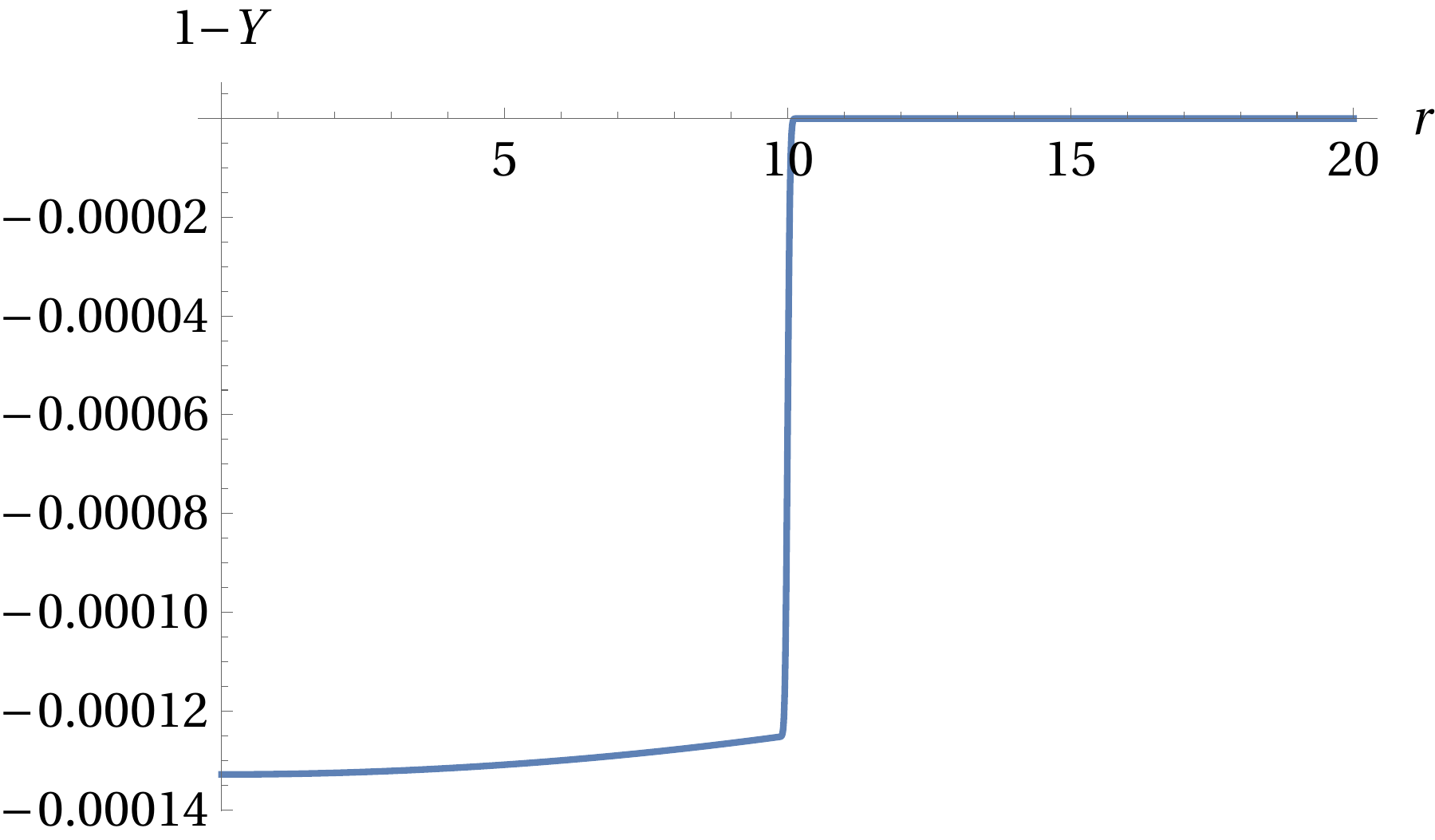}%
    }\\
  \subfloat[\label{fig:Ystaticb}]{%
    \includegraphics[width=\columnwidth]{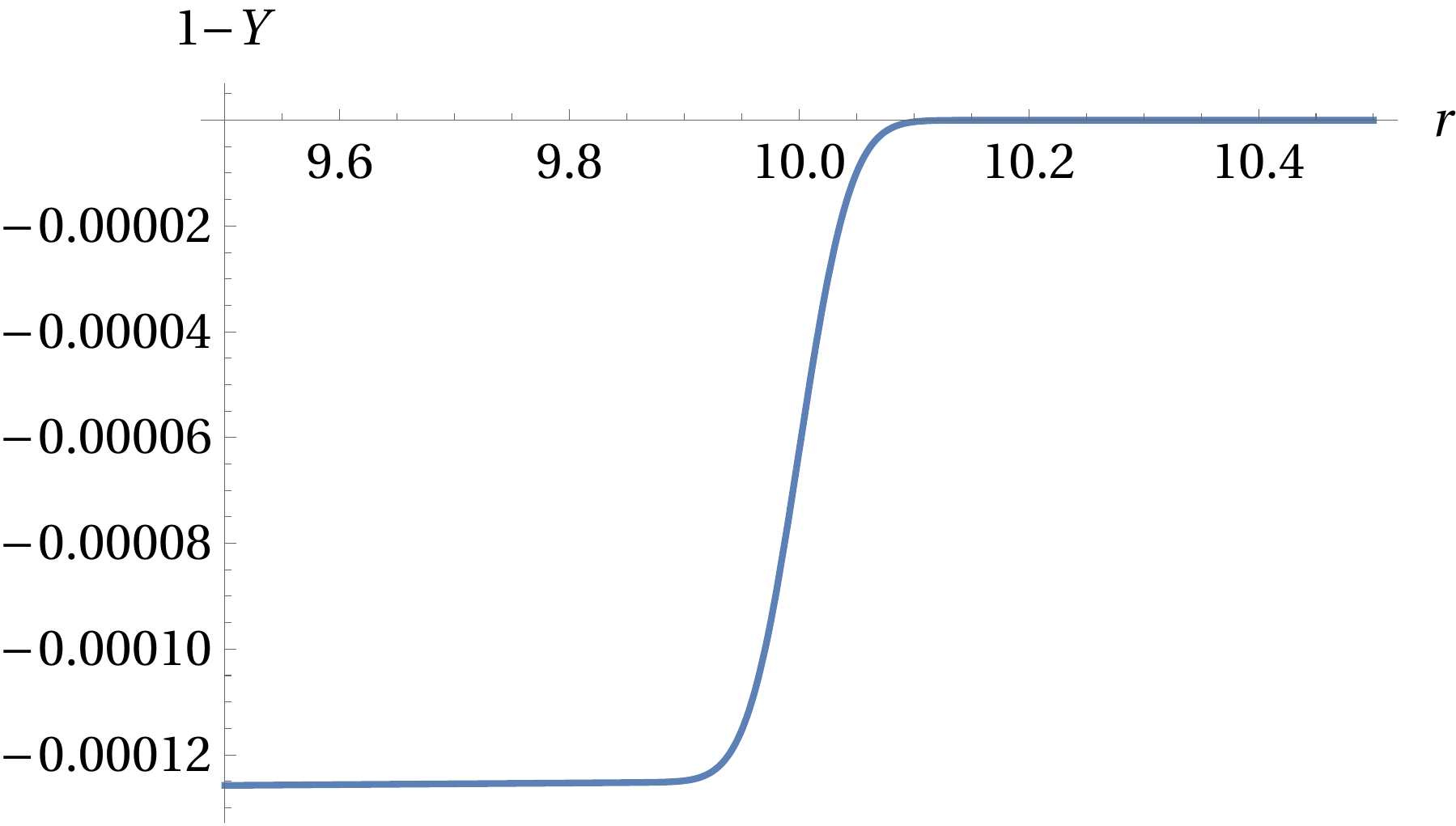}%
    }
  \caption{
    Differences between Jordan and Einstein frame profiles for the quantity $Y$ in terms of the Einstein frame areal radius $r$. (a) is the profile in the domain $[0,20]$ and (b) is the profile in the domain $[9.5,10.5]$.
  }
  \label{fig:Ystatic}
\end{figure}

Finally, we plot the quantity $Y$ in Fig. \ref{fig:Ystatic}, which parametrizes the differences between the Einstein and Jordan frame metrics. The results obtained here are solutions to the gravitational equation \eqref{GCA-GEN-GFE} and its divergence in the Einstein frame, which supplies the continuity equation used to derive the TOV equation. In the Einstein frame, there is nothing particularly pathological about the solution; dynamically, our result may be thought of as a solution to the Einstein equation for a fluid with a modified equation of state.
On the other hand, the results here illustrate the earlier observation that discontinuities in the matter density profile lead to discontinuities in the Jordan frame metric. Though one might expect sharp gradients to generate a strong backreaction effect on the matter, any backreaction has already been accounted for since we have obtained a solution to the full set of nonlinear equations.

\begin{figure}[!htp]
  \subfloat[\label{fig:Veff}]{%
    \includegraphics[width=\columnwidth]{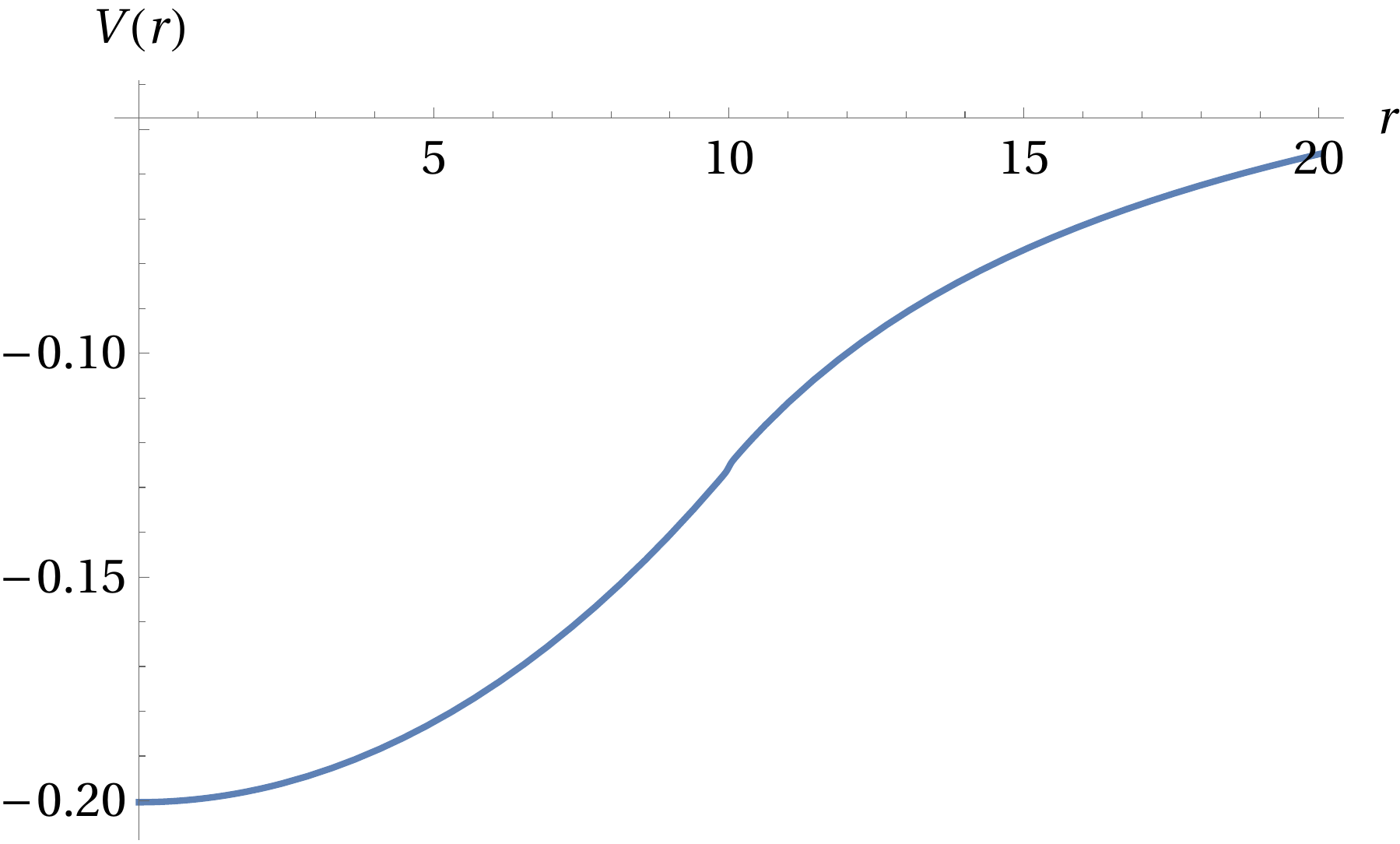}%
    }\\
  \subfloat[\label{fig:Feff}]{%
    \includegraphics[width=\columnwidth]{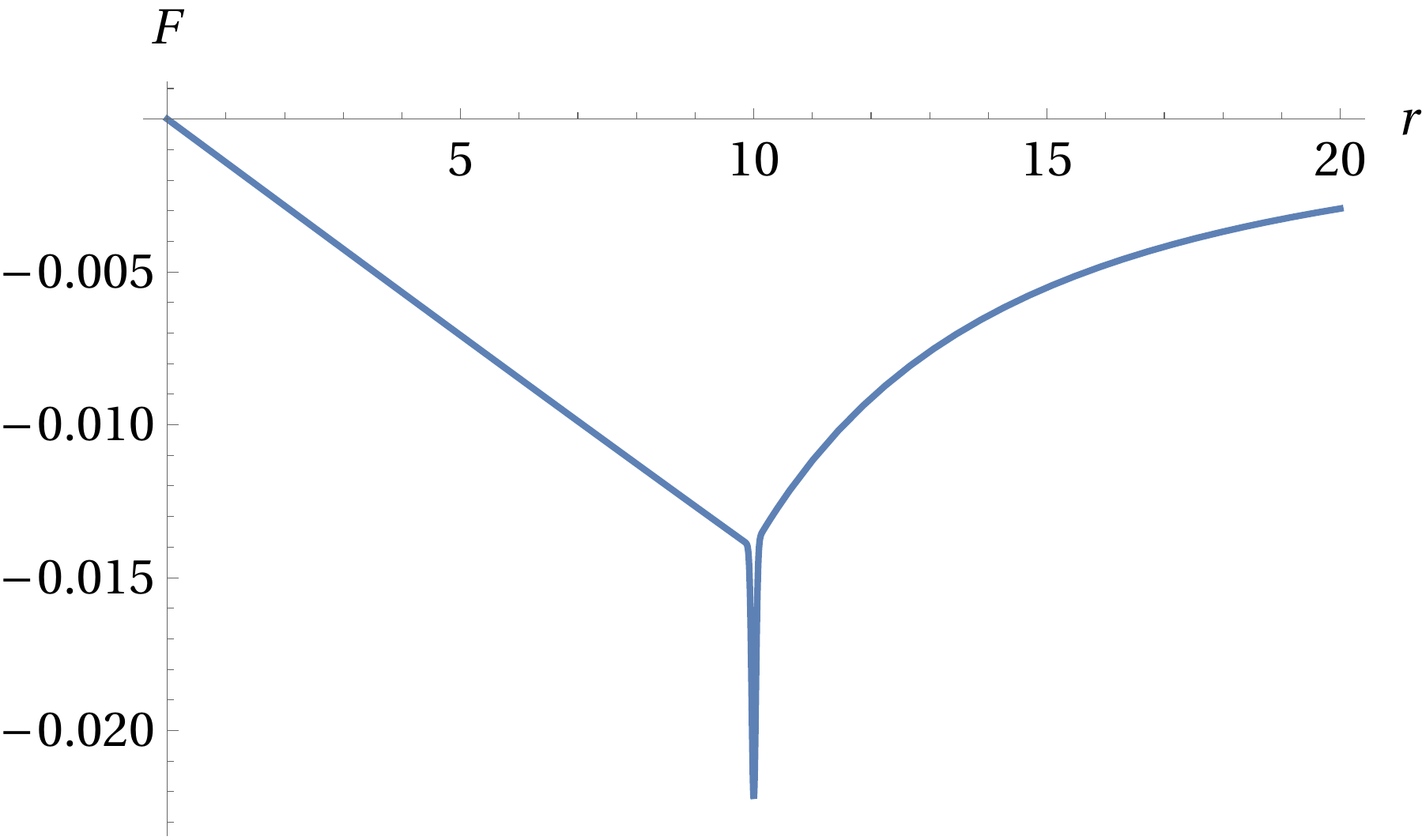}%
    }
  \caption{
    In (a), the effective potential $V(r)$ \eqref{EffectivePotential} for geodesics is displayed, with the parameter choices $R_0=10$, $\sigma=1/10$, $m_0=1$, and $q=-2\pi/3$, with $e=1$ and $l=0$ for the respective energy and angular momentum parameters. In (b), the effective force per unit mass $F$ is displayed. These plots are displayed in terms of the Einstein frame areal radius $r$, since it differs mildly from the proper radius $\textbf{r}$ for our parameter choices (one can verify that $Y \sqrt{h(r)} \approx 1$), so these plots still capture the qualitative behavior of the effective potential and force.
  }
  \label{fig:VFeff}
\end{figure}

On the other hand, one might expect geodesics in the Jordan frame metric to experience large accelerations in the presence of large gradients in the matter distribution. However, large accelerations do not necessarily lead to strong effects since such accelerations occur only in a very narrow region and all relevant physical quantities integrated over the thin layer are finite. To see this, we consider the behavior of geodesics in the line element \eqref{SpLE}, which are characterized by the invariants
\begin{equation}\label{Geodesics-ConservedQuantities}
\begin{aligned}
l &= \mathfrak{g}_{\mu 3} \frac{dx^\mu}{d\tau} = r^2 \, \dot{\phi}, \\
e &= \mathfrak{g}_{\mu 0} \frac{dx^\mu}{d\tau} = f \, \dot{t} ,
\end{aligned}
\end{equation}

\noindent where $l$ corresponds to an angular Killing vector, and $e$ corresponds to a timelike Killing vector. The overdot denotes the derivative with respect to the proper time $\tau$. One may rewrite the equation for $e$ as (assuming geodesics in the equatorial plane)
\begin{equation}\label{EnergyInvariant}
\frac{1}{2}\dot{\textbf{r}}^2 + V(r(\textbf{r})) = 0 ,
\end{equation}
\begin{equation}\label{EffectivePotential}
V(r):=\frac{1}{2} \left\{1-\frac{e^2 f(r)}{16 -  \left[24 - \left(f(r)^2+8\right) Y \right] Y}+\frac{l^2}{Y^2 r^2}\right\} ,
\end{equation}

\noindent where $\dot{\textbf{r}} = Y \sqrt{h(r)} \, \dot{r}$ is the time derivative with respect to proper radius $\textbf{r}$ in the Jordan frame [with $r(\textbf{r})$ defined implicitly by the differential expression $d\textbf{r}=Y \sqrt{h(r)} \, dr$]. From the above, one can obtain an effective radial force per unit mass $F=-Y^{-1} h(r)^{-1/2}\,\partial_r V(r)$. While the large gradients in $Y(r)$ generate spikes in the profile for the effective force per unit mass $F$, the jump in the effective potential is proportional to the jump in $Y$; this indicates that for particles crossing the strong gradient region, a small jump in $Y$ will correspond to a small change in energy, despite the large forces involved. This behavior is illustrated in Fig. \ref{fig:VFeff} for radial geodesics in our numerical solution of the TOV equation. In the plots, we see that for negative $q$, we see an increase in magnitude for the effective force (for positive $q$, the magnitude decreases for a range of values for the gradients), and is directed toward $r=0$. However, the effective potential only exhibits a modest jump near $r=10$.

It is perhaps appropriate to consider more realistic parameter choices. One can expand the Jordan frame metric [given by Eqs. \eqref{GCA-YSol} and \eqref{GCA-TransformedJordanMetric}] to leading order in $q \hat\rho_0$,
\begin{equation}\label{GCA-TransformedJordanMetricExpand}
\mathfrak{g}_{\mu \nu} = g_{\mu \nu} - q \hat \rho_0 \frac{(\hat \rho + \hat p)}{\hat \rho_0} \left[\frac{1}{2}g_{\mu \nu} + 2 u_\mu \, u_\nu \right] + \mathcal{O}([q \hat \rho_0]^2).
\end{equation}

\noindent The highest-energy densities probed to date in particle accelerators are 1-2 orders of magnitude higher than those in neutron stars; a density scale $1/|q|$ two orders of magnitude higher than the neutron star density ($\sim 0.4~{\rm GeV/fm^{3}}$) is consistent with collider data \cite{Andronic:2014zha,CMS:2012krf}. If this is the case, then a neutron star with a sharp boundary will contribute a rather large discontinuity of order $|q (\hat \rho + \hat p)| \sim 10^{-2}$ in the metric, which would lead to observable effects---one might, for instance, expect an additional heating (assuming $q<0$) of neutron star crusts from particles which pass through the crust on the order of $10^{-2}$ times the particle rest mass. For such large values of $q$, one might expect such effects to be observable in astrophysical phenomena, the neutrino cooling of neutron stars, for instance \cite{Brown:2017gxd,Lattimer:1991ib}. The absence of such an effect can place strong constraints on the parameter $q$.

%=======================================================================
%		SHARP GRADIENTS: DYNAMICAL CASE
%=======================================================================
\subsection{Dynamical case}
To understand the behavior of sharp density gradients in a dynamical situation, we consider the problem of pressureless collapse, like that of Oppenheimer-Snyder. It is appropriate to consider the problem in Painlev{\'e}-Gullstrand like coordinates, which are for the (exterior) Schwarzschild metric adapted to a class of observers free falling from rest at infinity. We consider a line element of the form
\begin{equation} \label{PGLE}
    ds^2 =  -\left(\alpha^2 - \beta^2\right) dt^2 
            + 2 \, \beta \, dt \, dr 
            + dr^2 +  r^2 \, d\Omega^2 .
\end{equation}

\noindent In the Einstein frame, the lowered index fluid four-velocity is parametrized as:
\begin{equation} \label{UL}
    u_{\cdot} = \left(u_t,v,0,0\right),
\end{equation}

\noindent where $u_t$ is determined by $u_\mu u^\mu = -1$, and $v=v(t,r)$ provides a measure of the radial velocity of the fluid relative to the unit vector normal to constant $t$ surfaces, which in the original Painlev{\'e}-Gullstrand metric corresponds to free-falling observers from rest at infinity (the four-velocity which are recovered when $v \rightarrow 0$). To simplify the analysis, we consider a parametrization in which the initial velocity profile vanishes; in particular, we set $v(0,r)=0$. The Einstein equations then yield the following constraint equations for the metric components at $t=0$
\begin{equation} \label{EFE-PGc}
    \begin{aligned}
        \partial_r\alpha  &= 0, \\
        \partial_r\beta  &= - \frac{\beta}{2 r} + \kappa r \frac{\alpha^2 \, \rho}{2 \beta} ,
    \end{aligned}
\end{equation}

\noindent so that $\alpha$ becomes independent of $r$ for the specified initial data. The constraint equations provide an interpretation for $v=v(t,r)$. If $\partial_r\alpha=0$ and $v(0,r)=0$, the acceleration vector $a^\mu = u^\sigma \nabla_\sigma u^\mu$ takes the form
\begin{equation} \label{EFE-Acc}
    a^\cdot = \left(0,\dot{v}/\alpha,0,0\right).
\end{equation}

\noindent For the initial data choice $v(0,r)=0$, the time derivative of the velocity $\dot{v}$ determines the radial acceleration of the fluid.

The evolution equations take the form
\begin{align}
        \dot{\alpha} &= \frac{\alpha}{2\beta} \left[2\dot{\beta}-\kappa \alpha^2 r \left(\rho + p\right)\right], \label{EFE-PG1}\\
 \partial_r \dot{\beta} &=
            \frac{\dot{\beta}}{4} \left[\frac{2 \kappa \alpha^2 r \rho}{\beta^2}-\frac{2}{r}\right]
            -\frac{\left(\kappa \alpha^2 r^2 p - \beta^2\right) \left(\kappa \alpha^2 r^2 \rho - 3 \beta^2\right)}{4 r^2 \beta^2} \nonumber\\
            & \quad 
            +\beta \, \partial^2_r\beta. \label{EFE-PG2}
\end{align}

\noindent Note that Eq. \eqref{EFE-PG2} does not determine $\dot{\beta}$ directly, but is a differential equation for $\dot{\beta}$ dependent on boundary conditions.

The conservation of the energy-momentum tensor $\nabla_\mu T^{\mu \nu} = 0$ yields the time derivative of the fluid density
\begin{equation} \label{DivT_Rhodot}
    \begin{aligned}
        \partial_t{\rho} - \beta \partial_r \rho = \frac{(\rho + p) \left(\kappa  r^2 \alpha^2 \rho+3 \beta^2\right)}{2 r \beta}.
    \end{aligned}
\end{equation}

\noindent For the unit vector $n^\mu$ normal to constant $t$ surfaces, one can construct the operator $\alpha n^\mu \partial_\mu = \partial_t - \beta \partial_r$; the left-hand side of Eq. \eqref{DivT_Rhodot} is proportional to the derivative of $\rho$ along $n^\mu$. The conservation of the energy-momentum tensor $\nabla_\mu T^{\mu \nu} = 0$ also yields the time derivative of the radial velocity
\begin{equation} \label{DivT_vdot}
    \dot{v} = -\frac{\alpha}{\rho + p} \partial_r p.
\end{equation}

\noindent We see here that the presence of a sharp density gradient in the effective pressure generates large accelerations in the radial velocity. From Eqs. \eqref{GCA-YSol}, \eqref{GCA-ExpAdetYform}, and \eqref{GCA-ExpTmnEffDecompDensityPressure}, one can obtain the following expression in terms of the Jordan frame fluid quantities $\hat \rho$ and $\hat p$
\begin{equation}\label{DivT-Deriv}
    \begin{aligned}
    \partial_r p = 
        |A| Y
        \left[\frac{3 q ( \hat \rho + \hat  p ) \partial_r \hat \rho }{4 (1+q \hat \rho)}
        +
        \frac{(1+q \hat \rho) \partial_r \hat p}{1 - q \hat p} \right],
    \end{aligned}
\end{equation}

\noindent and it follows that for finite $|q|$ and densities below the critical density $|q \hat\rho|=1$, large gradients in the Jordan frame density will generate large gradients in the Einstein frame pressure $p$. Large gradients in the density and pressure then yield large values for $\dot{v}$ in the vicinity of the gradient. From Eq. \eqref{EFE-Acc}, it follows that large pressure and density gradients in the Jordan frame generate large accelerations in the Einstein frame.

In the case of a pressureless dust fluid, one has
\begin{equation}\label{DivT-Gradrhodust}
    \begin{aligned}
    \partial_r p = 
        |A| Y \frac{3 q \hat \rho \partial_r \hat \rho }{4 (1+q \hat \rho)}.
    \end{aligned}
\end{equation}

\noindent If $|A|Y \hat\rho>0$ (assuming $|q \hat\rho| \ll 1$), the relative sign between the density and pressure gradients, and consequently the sign of $\dot{v}$, are determined by the sign of $q$. For negative $q$, $\dot{v}$ will have the same sign as the density gradient, and for positive $q$, $\dot{v}$ will have the opposite sign as the density gradient. For a spherically symmetric compact matter distribution with a monotonically decreasing density, this suggests a compressive force for $q<0$, which becomes strong near sharp boundaries. If the matter density is not monotonic in the radius, one can have nontrivial behavior for $q<0$. A thick shell of spherically symmetric dust will, for instance, become compressed into a thin shell, and dips in an otherwise uniform dust density profile will widen. This behavior also applies near the critical density for negative $q$, even when including pressure in the Jordan frame; in the limit $|q \rho|\rightarrow 1$, the density gradient dominates in Eq. \eqref{DivT-Deriv} (it is not divergent, since $|A|$ contributes a factor $(1+q \hat\rho)$). On the other hand, for $|q \hat\rho| \ll 1$, the pressure gradient in Eq. \eqref{DivT-Deriv} dominates, so for realistic matter models well below the critical density, the aforementioned effects are overwhelmed by pressure gradients.

%=======================================================================

%-----------------------------------------------------------------------
%-----------------------------------
%-----------------
%--------
%---
%-
%
%
%-
%---
%--------
%-----------------
%-----------------------------------
%-----------------------------------------------------------------------
%=======================================================================
%-----------------------------------------------------------------------
%
%		FINAL REMARKS
%
%-----------------------------------------------------------------------
%=======================================================================

\section{Final Remarks}\label{sec:Conc}

In this paper, we have developed the formalism to treat singular hypersurfaces and junction conditions in generalized coupling theories. In particular, junction conditions were obtained with a variational approach. Such an approach facilitates the derivation of the junction conditions and clarifies the dependence of the junction conditions on the assumed properties of the field at the junction surface. We have argued that the Einstein frame is fundamental, and with that viewpoint in mind, we employed the strategy of first deriving the junction conditions in the Einstein frame, then translating them into Jordan frame variables. A primary motivation for the development of the formalism stems from the observation (which we pointed out in \cite{Feng:2020lhp}) that the mixing of the spacetime and matter degrees of freedom in the Jordan frame can lead to spurious discontinuities in the metric---in particular, discontinuities in the energy-momentum tensor lead to discontinuities in the Jordan frame metric in GCTs. Nevertheless, we found that as long as the Einstein frame metric is continuous (taking the view that the Einstein frame is fundamental), the dynamics for the matter and auxiliary fields can be solved consistently.

The formalism we developed for junctions has been applied to the case of spherically symmetric matter distributions containing sharp gradients in the context of the MEMe models. In particular, we solved the TOV equation and, using Painlev{\'e}-Gullstrand like coordinates, we studied Oppenheimer-Snyder collapse.

In the first case, we found that discontinuities in the matter distribution only generate singularities in the Jordan frame picture, whereas, in the Einstein frame one, all equations are regular. Indeed in this case one can conclude that in a neutron star, a sharp boundary would lead to observable effects, e.g., expect an additional heating of neutron star crusts from incident particles on the order of $10^{-2}$ times the rest mass of the incident particles if the absolute value of the parameter $q$ is as large as the maximum value allowed by the current experimental upper bound. These effects are in principle measurable (in observations of neutrino cooling of neutron stars, for instance) and could lead to tighter constraints on the parameter $q$. 

In the dynamical case, instead, we find that during the initial stages of Oppenheimer-Snyder collapse, the matter distribution is subject to an effective force that depends on the matter distribution other than the value of the parameter $q$. This force tends to compress a matter distribution with a monotonically decreasing density profile, particularly near the boundary. This result suggests some possible interesting differences in the dynamics of gravitational collapse.

If the density is not monotonic, the net effect is to accentuate over-densities and under-densities in the matter distribution when one is near the critical density. This result has consequences at cosmological level. In particular, since in the matter era the density of matter is far from the critical density, cosmological structure formation in the context of MEMe does not differ significantly from the one in GR. This result, in turn, would imply that the MEMe model is not able to account for cosmological dark matter. Further analysis of this problem, and in particular a complete analysis of the formation of structures in the context of the MEMe model, will lead to further insight into the actual mechanism of structure formation in this context.

%=======================================================================

%-----------------------------------------------------------------------
%-----------------------------------
%-----------------
%--------
%---
%-
%
%
%-
%---
%--------
%-----------------
%-----------------------------------
%-----------------------------------------------------------------------
%=======================================================================

%=======================================================================
%		ACKNOWLEDGMENTS
%=======================================================================

\begin{acknowledgments}
J.C.F. is grateful to the Department of Mechanical, Energy, Management and Transportation Engineering (DIME) at the University of Genoa for hosting a visit during which part of this work was performed, and acknowledges support from Funda\c{c}\~{a}o para a Ci\^{e}ncia e a Tecnologia Grants No. PTDC/MAT-APL/30043/2017 and No. UIDB/00099/2020. The work of S.M. was supported in part by Japan Society for the Promotion of Science Grants-in-Aid for Scientific Research No.~17H02890, and No.~17H06359 and by World Premier International Research Center Initiative, The Ministry of Education, Culture, Sports, Science and Technology, Japan. Some of the calculations were performed using the xAct package \cite{xActbib} for \textit{Mathematica}.
\end{acknowledgments}

\appendix*

%=======================================================================
%-----------------------------------------------------------------------
%
%		APPENDIX:
%
%-----------------------------------------------------------------------
%=======================================================================
\section{VARIATION OF THE GRAVITATIONAL ACTION} \label{VarGrav}
To obtain the variation of the gravitational action, one should write it in the form,
\begin{equation} \label{GC:ActionGrav2}
   S_{\rm G}[{g}^{\cdot \cdot}] = \frac{1}{2 \kappa} \left[ \int_U d^4 x \sqrt{|g|} \left( R - 2 \Lambda \right) + 2 \varepsilon \int_{\Sigma} d^3y \sqrt{|q|} \, K \right] ,
\end{equation}

\noindent where the surface $\Sigma = \partial U$ is a limiting surface of a foliation, so that the vector field $n^\mu$ can be defined in a neighborhood of $\Sigma$. The full (Weiss \cite{Feng:2017ygy,Feng:2021lfa}) variation of Eq. \eqref{GC:ActionGrav2}, including displacements of the boundary, takes the form
\begin{equation} \label{GC:ActionGrav2bVar}
   \begin{aligned}
   \delta S_{\rm G}= \frac{1}{2 \kappa} 
   \biggl[ 
      &
      \int_U d^4 x \sqrt{|g|} \biggl\{ 
      (G_{\mu \nu} + \Lambda g_{\mu \nu}) \delta g^{\mu \nu}
      +
      g^{\mu \nu}
      \delta R_{\mu \nu} 
      \biggr\} \\
      & 
      +
      \varepsilon \, \int_{\Sigma} d^3y \sqrt{|q|} \biggl\{\left( R - 2 \Lambda \right) n_\mu \, \delta Z^\mu + 2 \, \Delta K \\
      & \qquad \qquad \qquad \qquad
      - K \, q_{\mu \nu} \, \Delta q^{\mu \nu} \biggr\}
   \biggr],
   \end{aligned}
\end{equation}

\noindent where $\delta Z^{\mu}$ is the variation of the embedding functions $Z^{\mu}(y)$ of the junction hypersurface~\cite{Mukohyama:2001pb}, $\Delta$ corresponds to the change in the quantity accounting for boundary displacements. It is not too difficult to show that (with the understanding that the raising and lowering of indices on the variation of connections occurs after the variation is performed)
\begin{equation} \label{GC:RicciVar}
   \begin{aligned}
      g^{\mu \nu} \delta R_{\mu \nu}
         &= {\nabla}_\mu \left(\delta{^{\mu \alpha}}{_{\nu \sigma}} \, g^{\sigma \beta} \delta {\Gamma}{^\nu}{_{\alpha \beta}}\right) \\
         &= {\nabla}_\mu \left(
            \delta {\Gamma}{^{\mu \sigma}}{_{\sigma}}
            -
            \delta {\Gamma}{^{\sigma \mu}}{_{\sigma}}
            \right), 
   \end{aligned}
\end{equation}

\noindent with
\begin{equation} \label{GKD}
  \begin{aligned}
  \delta{^{\mu \alpha}}{_{\nu \sigma}}
      & := \delta{^\mu}{_\nu} \delta{^\alpha}{_\sigma} - \delta{^\mu}{_\sigma} \delta{^\alpha}{_\nu} .
\end{aligned}
\end{equation}

\noindent The variation of the action becomes
\begin{equation} \label{GC:ActionGrav2bVar2}
   \begin{aligned}
   \delta S_{\rm G} = \frac{1}{2 \kappa} 
   \biggl[ 
      &
      \int_U d^4 x \sqrt{|g|}
      (G_{\mu \nu} + \Lambda g_{\mu \nu}) \delta g^{\mu \nu}+ \\
      &
      \varepsilon \, \int_{\Sigma} d^3y \sqrt{|q|} \biggl\{\left(\delta{^{\mu \alpha}}{_{\nu \sigma}} \, g^{\sigma \beta} \delta {\Gamma}{^\nu}{_{\alpha \beta}}\right)n_\mu + \\
      &
      \left( R - 2 \Lambda \right) n_\mu \, \delta Z^\mu + 2 \Delta K - K q_{\mu \nu} \Delta q^{\mu \nu} \biggr\}
   \biggr].
   \end{aligned}
\end{equation}

\noindent The change in the connection, accounting for boundary displacements, takes the form
\begin{equation} \label{ChangeConnection}
   \Delta \Gamma{^\sigma}_{\mu \nu} = \delta \Gamma{^\sigma}_{\mu \nu} + \pounds_{\delta Z} \Gamma{^\sigma}_{\mu \nu}.
\end{equation}

\noindent The Lie derivative of the connection takes the form
\begin{equation} \label{LDConnection}
   \pounds_{\delta Z} \Gamma{^\sigma}_{\mu \nu} = {\nabla}_\mu {\nabla}_\nu \delta Z^\gamma + \delta Z^\sigma {R}{^\gamma}_{\nu \sigma \mu},
\end{equation}

\noindent where
\begin{equation} \label{LDC2}
   \delta{^{\mu \alpha}}{_{\sigma \tau}} \, g^{\tau \beta} \, \pounds_\xi {\Gamma}^\sigma{_{\alpha \beta}}
   = 2 {R}{^\mu}{_\sigma} \delta Z^\sigma - J^\mu,
\end{equation}

\noindent with the current
\begin{equation} \label{Current}
   J^\mu := {\nabla}_\sigma \left( {\nabla}^\mu \delta Z^\sigma - {\nabla}^\sigma \delta Z^\mu \right).
\end{equation}

\noindent The variation of the action takes the form
\begin{equation} \label{GC:ActionGrav2bVar1}
   \begin{aligned}
   \delta S_{\rm G}= \frac{1}{2 \kappa} 
   \biggl[ 
      &
      \int_U d^4 x \sqrt{|g|}
      (G_{\mu \nu} + \Lambda g_{\mu \nu}) \delta g^{\mu \nu} \\
      &
      - \varepsilon \int_{\Sigma} d^3y \sqrt{|q|} n_\mu \biggl\{2\left( G{^\mu}{_\nu} + \Lambda \delta{^\mu}{_\nu} \right) \delta Z^\nu - J^\mu \biggr\}\\
      &
      +
      \varepsilon \int_{\Sigma} d^3y \sqrt{|q|} \biggl\{\left(\Delta {\Gamma}{^{\mu \sigma}}{_{\sigma}} - \Delta {\Gamma}{^{\sigma \mu}}{_{\sigma}}\right)n_\mu + 2 \Delta K \\
      & \qquad \qquad \qquad \qquad
      - K \, q_{\mu \nu} \, \Delta q^{\mu \nu} \biggr\}
   \biggr].
   \end{aligned}
\end{equation}

\noindent To deal with the last set of boundary terms, it is perhaps appropriate to work out the variation of the extrinsic curvature, which takes the form
\begin{equation} \label{ExtrinsicCurvature}
   \begin{aligned}
   K_{\mu \nu}
       & :=  q{^\alpha}{_\mu} \, q{^\beta}{_\nu} \, \nabla_\alpha n_\beta \\
       & \,= q{^\alpha}{_\mu} \, \nabla_\alpha n_\nu \\
       & \,= \nabla_\mu n_\nu - \varepsilon \, n_\mu \, a_\nu.
\end{aligned}
\end{equation}

\noindent The mean curvature is given by the trace $K = g^{\mu \nu} \, K_{\mu \nu} = \nabla_\sigma n^\sigma$, where $n^\sigma \, a_\sigma = 0$. The variation of the extrinsic curvature may be
\begin{equation} \label{ExtrinsicCurvatureVar}
   \begin{aligned}
   \delta K_{\mu \nu}
       & \,= \delta q{^\alpha}{_\mu} \, \nabla_\alpha n_\nu + q{^\alpha}{_\mu} \, \nabla_\alpha \delta n_\nu - q{^\alpha}{_\mu} \, \delta \Gamma{^\sigma}_{\alpha \nu} \, n_\sigma \\
       & \,= - \varepsilon \, \delta n{^\alpha} n{_\mu} \, \nabla_\alpha n_\nu - \varepsilon \, \delta n{_\mu} \, a_\nu + q{^\alpha}{_\mu} \, \nabla_\alpha \delta n_\nu \\
       & \quad \,
       - q{^\alpha}{_\mu} \, \delta \Gamma{^\sigma}_{\alpha \nu} \, n_\sigma.
\end{aligned}
\end{equation}

\noindent The variation of the mean curvature may be written in three different ways
\begin{equation} \label{MeanCurvatureVar}
   \begin{aligned}
   \Delta K
      & \,= \Delta q^{\mu \nu} \, K_{\mu \nu} + q^{\mu \nu} \, \Delta K_{\mu \nu} \\
      & \,= \Delta q^{\mu \nu} \, K_{\mu \nu} + q^{\mu \nu} \left( \nabla_\mu \Delta n_\nu - \varepsilon \, \Delta n_\mu \, a_\nu - n_\sigma \, \Delta\Gamma{^\sigma}_{\mu \nu} \right) \\
      & \,= \nabla_\sigma \Delta n^\sigma + \Delta \Gamma{^{\tau \sigma}}_{\tau} n_\sigma.
\end{aligned}
\end{equation}

\noindent Now if $n_\sigma \Delta n^\sigma = - \Delta n_\sigma n^\sigma$, one may write
\begin{equation} \label{DVarn}
   \begin{aligned}
   D_\sigma \Delta n^\sigma 
        & = \nabla_\sigma \Delta n^\sigma - \varepsilon n{^\sigma}n{_\tau} \nabla_\sigma \Delta n^\tau \\
    & = \nabla_\sigma \Delta n^\sigma + \varepsilon \, n{_\tau} \, n{^\sigma} \, n^\nu \, \Delta \Gamma{^\tau}_{\sigma \nu},
\end{aligned}
\end{equation}

\noindent where the last line makes use of the variation of the acceleration vector
\begin{equation} \label{AccelVar}
   \begin{aligned}
      \Delta a^\tau = \Delta n{^\sigma} \nabla_\sigma n^\tau + n{^\sigma} \nabla_\sigma \Delta n^\tau + n{^\sigma} \, n^\nu \, \Delta \Gamma{^\tau}_{\sigma \nu}.
\end{aligned}
\end{equation}

\noindent One may also note that
\begin{equation} \label{qdn}
   \begin{aligned}
      q^{\mu \nu} \, \nabla_\mu \Delta n_\nu 
      & = D_\mu \left(g^{\mu \nu} \Delta n_\nu \right).
\end{aligned}
\end{equation}

\noindent The variation of the mean curvature may then be written [adding the second two lines of Eq. \eqref{MeanCurvatureVar}]
\begin{equation} \label{MeanCurvatureVar2}
   \begin{aligned}
   2\Delta K
            & \,= \Delta q^{\mu \nu} \, K_{\mu \nu} - \varepsilon \, q^{\mu \nu} \, \Delta n_\mu \, a_\nu + q^{\mu \nu} \, \nabla_\mu \Delta n_\nu \\
            & \quad \, - q^{\mu \nu} \, n_\sigma \, \Delta\Gamma{^\sigma}_{\mu \nu} 
            +
            \nabla_\sigma \Delta n^\sigma + \Delta \Gamma{^{\tau \sigma}}_{\tau} n_\sigma \\
      & \,= \Delta q^{\mu \nu} \, K_{\mu \nu} - \varepsilon \, \Delta n_\mu \, a^\mu + D_\mu \left(\Delta n^\mu + g^{\mu \nu} \Delta n_\nu \right)\\
      & \quad \,
      +
      n_\sigma\left(\Delta \Gamma{^{\tau \sigma}}_{\tau} - \Delta \Gamma{^{\sigma \tau}}_{\tau} \right) .
\end{aligned}
\end{equation}

\noindent From which one may obtain the expression
\begin{equation} \label{GammaMinusGamma}
   \begin{aligned}
      n_\sigma\left( \Delta \Gamma{^{\sigma \tau}}_{\tau} - \Delta \Gamma{^{\tau \sigma}}_{\tau} \right)
      & \,= \Delta q^{\mu \nu} \, K_{\mu \nu} - \varepsilon \, \Delta n_\mu \, a^\mu \\
      & \quad \, + D_\mu \left(\Delta n^\mu + g^{\mu \nu} \Delta n_\nu \right) - 2\Delta K .
\end{aligned}
\end{equation}

\noindent The variation of the action becomes
\begin{equation} \label{GC:ActionGrav2bVar3}
   \begin{aligned}
   \delta S_{\rm G} = \frac{1}{2 \kappa} 
   \biggl[ 
      &
      \int_U d^4 x \sqrt{|g|}
      (G_{\mu \nu} + \Lambda g_{\mu \nu}) \delta g^{\mu \nu} \\
      & - \varepsilon \int_{\Sigma} d^3y \sqrt{|q|} \,n_\mu \left\{ 2\left( G{^\mu}{_\nu} + \Lambda \delta{^\mu}{_\nu} \right) \delta Z^\nu  - J^\mu \right\}\\
      &
      +
      \varepsilon \int_{\Sigma} d^3y \sqrt{|q|} \biggl\{\left(K_{\mu \nu} - K \, q_{\mu \nu} \right) \Delta q^{\mu \nu} \\
      & \qquad \qquad + D_\mu \left(\Delta n^\mu + g^{\mu \nu} \Delta n_\nu \right) - \varepsilon \, \Delta n_\mu \, a^\mu  \biggr\}
   \biggr].
   \end{aligned}
\end{equation}

\noindent One may choose the foliation such that $a^\mu = 0$, and noting that $\Delta n^\mu + g^{\mu \nu} \Delta n_\nu$ is a purely spatial vector under the condition [$\Delta (n_\mu n^\mu)=0$],
the $D_\mu$ term becomes a surface term which we may discard. For topologically simple (contractible or star-shaped) domains, the term involving $J^\mu$ can be converted into a bulk integral of the divergence $\nabla_\mu J^\mu = 0$, which identically vanishes. The variation reduces to
\begin{equation} \label{GC:ActionGrav2bVar4}
   \begin{aligned}
   \delta S_{\rm G} = \frac{1}{2 \kappa} 
   \biggl[ 
      &
      \int_U d^4 x \sqrt{|g|}
      (G_{\mu \nu} + \Lambda g_{\mu \nu}) \delta g^{\mu \nu} \\
      & - 2 \, \varepsilon \int_{\Sigma} d^3y \sqrt{|q|} \left(G{^\mu}{_\nu} + \Lambda  \delta{^\mu}{_\nu} \right) n_\mu \, \delta Z^\nu \\
      &
      + \varepsilon \int_{\Sigma} d^3y \sqrt{|q|} \left(K_{\mu \nu} - K \, q_{\mu \nu} \right) \Delta q^{\mu \nu} 
   \biggr].
   \end{aligned}
\end{equation}

%   \vfill

%=======================================================================
%
%		APPENDIX: SUBSECTION
%
%=======================================================================
%	\subsection{APPENDIXSUBSECTION}

%=======================================================================
%		BIBLIOGRAPHY
%=======================================================================

\bibliography{JuncGC}

%\pagebreak

%=======================================================================

%-----------------------------------------------------------------------
%-----------------------------------
%-----------------
%--------
%---
%-
%
%
%-
%---
%--------
%-----------------
%-----------------------------------
%-----------------------------------------------------------------------

%=======================================================================
%-----------------------------------------------------------------------
%
%		APPENDICES
%
%-----------------------------------------------------------------------

\end{document}